\documentclass[journal]{IEEEtran}
\usepackage{amsmath,amsthm}
\usepackage{amssymb,bm}
\usepackage{graphicx}
\usepackage{multirow}
\usepackage{subcaption} 
\usepackage{color,soul}
\DeclareMathOperator*{\argmax}{arg\,max}
\usepackage{enumitem}

\usepackage{xcolor}
\usepackage{framed}

\usepackage{algorithm}
\usepackage{algcompatible}

\newtheorem{Lemma}{Lemma}
\newtheorem{Prop}{Proposition}
\newtheorem{Theorem}{Theorem}

\newtheorem{Def}{Definition}
\newtheorem{Remark}{Remark}
\definecolor{shadecolor}{RGB}{220,220,220}

\newcommand{\rpm}{\raisebox{.2ex}{$\scriptstyle\pm$}}
\begin{document}

\title{Tensors, Learning, and `Kolmogorov Extension' for Finite-alphabet Random Vectors}
\vspace{-5mm}
\author{Nikos Kargas, Nicholas D. Sidiropoulos, \emph{Fellow, IEEE}, and Xiao Fu, \emph{Member, IEEE}

	\thanks{Original manuscript submitted to {\it IEEE Trans. on Signal Processing} November 30, 2017; revised April 24, 2018; accepted June 28, 2018. Supported in part by NSF IIS-1447788 and IIS-1704074. Conference version of part of this work appeared in \textit{Information Theory and Applications Workshop} 2017~\cite{KaSi2017}.}
	\thanks{N. Kargas is with the Dept. of ECE, Univ. of Minnesota, Minneapolis, MN 55455; N. D. Sidiropoulos is with the Dept. of ECE, Univ. of Virginia, Charlottesville, VA 22904; X. Fu is with the School of EE and CS, Oregon State University, Corvallis, OR 97330. Author e-mails: karga005@umn.edu, nikos@virginia.edu, xiao.fu@oregonstate.edu}
	}
	\vspace{-5mm}
\maketitle

\begin{abstract}
Estimating the joint probability mass function (PMF) of a set of random variables lies at the heart of statistical learning and signal processing. Without structural assumptions, such as modeling the variables as a Markov chain, tree, or other graphical model, joint PMF estimation is often considered mission impossible -- the number of unknowns grows exponentially with the number of variables. But who gives us the structural model? Is there a {\em generic}, `non-parametric' way to control joint PMF complexity without relying on {\em a priori} structural assumptions regarding the underlying probability model? Is it possible to {\em discover} the operational structure without biasing the analysis up front? What if we only observe random subsets of the variables, can we still reliably estimate the joint PMF of all? This paper shows, perhaps surprisingly, that if the joint PMF of any three variables can be estimated, then the joint PMF of all the variables can be provably recovered under relatively mild conditions. The result is reminiscent of Kolmogorov's extension theorem -- consistent specification of lower-dimensional distributions induces a unique probability measure for the entire process. The difference is that for processes of limited complexity (rank of the high-dimensional PMF) it is possible to obtain complete characterization from only three-dimensional distributions. In fact not all three-dimensional PMFs are needed; and under more stringent conditions even two-dimensional will do. Exploiting multilinear (tensor) algebra, this paper proves that such higher-dimensional PMF completion can be guaranteed -- several pertinent identifiability results are derived. It also provides a practical and efficient algorithm to carry out the recovery task. Judiciously designed simulations and real-data experiments on movie recommendation and data classification are presented to showcase the effectiveness of the approach.

\end{abstract}
\IEEEpeerreviewmaketitle
\section{Introduction}

Estimating a joint Probability Mass Function (PMF) of a set of random variables is of great interest in numerous applications in the fields of machine learning, data mining and signal processing.  In many cases, we are given partial observations and/or statistics of the data, i.e., incomplete data, marginalized lower-dimensional distributions, or lower-order moments of the data, and our goal is to estimate the missing data. If the full joint PMF of all variables of interest were known, this would have been a straightforward task.  A classical example is in recommender systems, where users rate only a small fraction of the total items (e.g., movies) and the objective is to make item recommendations to users according to predicted ratings. If the joint PMF of the item ratings is known, such recommendation is readily implementable based on the conditional expectation or mode of the unobserved ratings given the observed ratings. A closely related problem is top-$K$ recommendation, where the goal is to predict the $K$ items that a user is most likely to buy next. When the joint PMF of the items is known, it is easy to identify the $K$ items with the highest individual or joint (`bundle') conditional probability given the observed user ratings. Another example is data classification. If the joint PMF of the features and the label is known, then given a test sample it is easy to infer the label according to the Maximum a \textit{Posteriori} (MAP) principle.  In fact, the joint PMF can be used to infer any of the features (or subsets of them), which is useful in imputing incomplete information in surveys or databases.

Despite its importance in signal and data analytics, estimating the joint PMF is often considered mission impossible in general, if no structure or relationship between the variables (e.g., a tree structure or a Markovian structure) can be assumed.
This is true even when the problem size is merely moderate. The reason is that the number of unknown parameters is exponential in the number of variables. Consider a simple scenario of $10$ variables taking $10$ distinct values each. The number of parameters we need to estimate in this case is $10^{10}$. The `naive' approach for joint PMF estimation is counting the occurences of the joint variable realizations. In practice, however, when dealing with even moderately large sets of random variables, the probability of encountering any particular realization is very low. Therefore, only a small portion of the empirical distribution will be non-zero given a reasonable amount of data samples -- this makes the approach very inaccurate.

\begin{figure}
	\centering
	\includegraphics[width=0.6\linewidth]{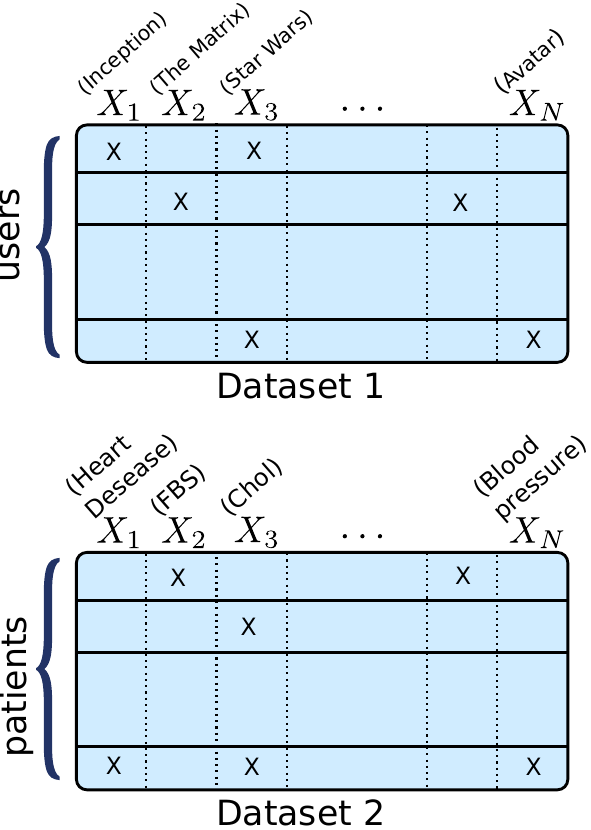}
	\caption{Applications of joint PMF estimation. Top: recommender systems: given partially observed ratings of a user on movies, we would like to infer the unobserved ratings.
		Bottom: classification problems: given medical features of people, we would like to infer if a person has heart disease.}
	\label{fig:intro}
\end{figure}

In many applications, different workarounds have been proposed to circumvent this sample complexity problem. 
For example, in recommender systems, instead of trying to estimate the joint PMF of the ratings (which would be the estimation-theoretic gold standard), the most popular approach is based on low-rank matrix completion~\cite{Koren2009,JaNeSa2013,MnSa2008}.
The idea is that the users can be roughly clustered into several types, and users of the same type would rate different movies similarly. Consequently, the user-rating matrix is approximately low rank and this is used as prior information to infer the missing ratings. In classification, parsimonious function approximations are employed to model the relationship (or the conditional probability function) between the features and the label. Successful methods that fall into this category are support vector machines (linear function approximation), logistic regression (log-linear function approximation) and more recently kernels and neural networks (nonlinear function approximation)~\cite{Bi2006}.

The above mentioned methods are nice and elegant, have triggered a tremendous amount of theoretical research and practical applications, and have been successful in many ways. However, these workarounds have not yet answered our question of interest: Can we ever reliably estimate the joint PMF of variables given limited data? This question is very well-motivated in practice, since knowledge of the joint PMF is indeed the gold standard: it enables optimal estimation under a variety of well-established criteria, such as mean-square error and minimum probability of error or Bayes risk. Knowing the joint PMF can facilitate a large variety of applications including recommender systems and classification in a unified and statistically optimal way, instead of resorting to often ad-hoc modeling tools. 

This paper shows, perhaps surprisingly, that if the joint PMF of any three variables can be estimated, then the joint PMF of all the variables can be provably recovered under relatively mild conditions. The result is reminiscent of Kolmogorov's extension theorem -- consistent specification of {lower-dimensional} distributions induces a unique probability measure for the entire process. The difference is that for processes of limited complexity (rank of the {high-dimensional} PMF) it is possible to obtain complete characterization from only {three-dimensional} distributions. In fact not all { three-dimensional} PMFs are needed; and under more stringent conditions even {two-dimensional} will do. The rank condition on the { high-dimensional} joint PMF has an interesting interpretation: loosely speaking, it means that  the random variables are `reasonably (in)dependent'. This makes sense, because estimation problems involving fully independent or fully dependent regressors and unknowns are contrived -- it is the middle ground that is interesting. It is also important to note that the marginal PMFs of triples can be reliably estimated at far smaller sample complexity than the joint PMF of all variables. For example, for user-movie ratings, the marginal PMF of three given variables (movies) can be estimated by counting the co-occurrences of the given ratings (values of the variables) of the three given movies; but no user can rate {\em all} movies. 

\smallskip

\noindent
{\bf Contributions} Our specific contributions are as follows:

\noindent
$\bullet$ We propose a novel framework for joint PMF estimation given limited and possibly very incomplete data samples. Our method is based on a nice and delicate connection between the Canonical Polyadic Decomposition (CPD)~\cite{CaCha1970,Har1970} and the naive Bayes model. The CPD model, sometimes referred to as the Parallel Factor Analysis (PARAFAC) model, is a popular analytical tool from multiway linear algebra. The CPD model has been used to model and analyze tensor data (data with more than two indices) in signal processing and machine learning, and it has found many successful applications, such as speech separation \cite{NiSi2010}, blind CDMA detection~\cite{SiGiBro2000}, array processing \cite{SiBroGi2000}, spectrum sensing and unmixing in cognitive radio~\cite{FuSiTra2015}, topic modeling \cite{AnGeHsu2014b}, and community detection \cite{AnGeHsu2014a} -- see the recent overview paper in~\cite{SiDeFu2017}.  Nevertheless, CPD has never been considered as a statistical learning tool for recovering a \emph{general} joint PMF and our work is the first to establish the exciting connection \footnote{There are works that considered using CPD to model a joint PMF for some specific problems~\cite{AnGeHsu2014b}. However, these works rely on specific physical interpretation of the associated model, which is sharply different to our setup -- in which we employ the CPD model to explain a general joint PMF without assuming any physical model.}.

\noindent
$\bullet$  
We present detailed identifiability analysis of the proposed approach.
We first show that, any joint PMF can be represented by a naive Bayes model with a finite-alphabet latent variable -- and the size of the latent alphabet (which happens to be the {\em rank} of the joint PMF tensor, as we will see) is bounded by a function of the alphabet sizes of the (possibly intermittently) observed variables. We further show that, if the latent alphabet size is under a certain threshold, then the joint PMF of {\it an arbitrary number} of random variables can be identified from { three-dimensional} marginal distributions. We prove this identifiability result by relating the joint PMF and marginal PMFs to the CPD model, which is known for its uniqueness even when the tensor rank is much larger than its outer dimensions.

\noindent
$\bullet$ In addition to the novel formulation and identifiability results, we also propose an easily implementable joint PMF recovery algorithm. Our identification criterion can be considered as a coupled simplex-constrained tensor factorization problem, and we propose a very efficient alternating optimization-based algorithm to handle it. To deal with the probability simplex constraints that arise for PMF estimation, the celebrated Alternating Direction Method of Multipliers (ADMM) algorithm is employed, resulting in lightweight iterations.  Judiciously designed simulations and real experiments on movie recommendation and classification tasks are used to showcase the effectiveness of the approach.

Preliminary version of part of this work appeared at ITA 2017~\cite{KaSi2017}. This journal version includes new and stronger identifiability theorems and interpretations, detailed analysis of the theorems, and insightful experiments on a number of real datasets.

\subsection{Notation} 
Bold, lowercase and uppercase letters denote vectors and matrices respectively. Bold, underlined, uppercase letters denote $N$-way ($N \geq 3$) tensors. Uppercase (lowercase) letters denote scalar random variables (realizations thereof, respectively). The outer product of $N$ vectors is a $N$-way tensor with elements $(\mathbf{a}_1 \circ \mathbf{a}_2 \cdots \circ \mathbf{a}_N)(i_1,i_2,\ldots,i_N) = \mathbf{a}_1(i_1)\mathbf{a}_2(i_2)\cdots \mathbf{a}_N(i_N)$. The Kronecker product of matrices $\mathbf{A}$  and $\mathbf{B}$ is denoted as $\mathbf{A} \otimes \mathbf{B}$. The Khatri-Rao (column-wise Kronecker) product of matrices $\mathbf{A}$ and $\mathbf{B}$ is denoted as $\mathbf{A} \odot \mathbf{B}$. The Hadamard (element-wise) product of matrices $\mathbf{A}$ and $\mathbf{B}$ is denoted as  $\mathbf{A} \circledast \mathbf{B}$. We define $\text{vec}(\underline{\mathbf{X}})$ the vector obtained by vertically stacking the elements of a tensor $\underline{\mathbf{X}}$ into a vector. Additionally, $\textrm{diag}(\mathbf{x}) \in \mathbb{R}^{I \times I}$ denotes the diagonal matrix with the elements of vector $\mathbf{x}  \in \mathbb{R}^{I}$ on its diagonal. The set of integers $\mathcal{S} = \{1,\ldots,N\}$ is denoted as $[N]$ and $|\mathcal{S}|$ denotes the cardinality of the set $\mathcal{S}$.
\begin{figure}[t]
	\centering
	\includegraphics[width=0.7\linewidth]{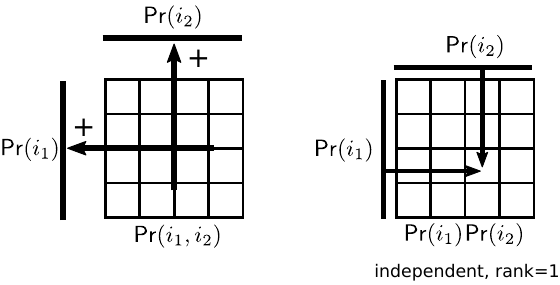}
	\caption{It is impossible to recover the joint PMF from {one-dimensional} marginals {without making strong assumptions}.}
	\label{fig:rank1}
\end{figure}

\section{Problem Statement}
\label{sec:pr_statement}
Consider a set of $N$ random variables, i.e., $\{X_n\}_{n=1}^N$. Assume that each $X_n$ can take $I_n$ discrete values and only the joint PMFs of variable triples, i.e., ${\sf Pr}(X_j=i_j,X_k=i_k,X_\ell=i_\ell)$'s, are available. Can we identify the joint PMF of $\{X_n\}_{n=1}^N$, i.e., ${\sf Pr}(X_1=i_1,\ldots,X_N=i_N) $, from the { three-dimensional} marginals? This question lies at the heart of statistical learning. To see this, consider a classification problem and  let $X_1\ldots,X_{N-1}$ represent the set of observed features, and $X_N$ the sought label. If ${\sf Pr}(X_1=i_1,\ldots,X_N=i_N) $ is known, then given a specific realization of the features, one can easily compute the posterior probability 
\[{\sf Pr}(i_N | i_1 \ldots, i_{N-1})=\frac{{\sf Pr}(i_1,\ldots,i_N)}{ \sum_{i_N=1}^{I_N}{\sf Pr}(i_1,\ldots,i_{N-1},i_N)},\]
and predict the label according the MAP principle (here ${\sf Pr}(i_1,\ldots,i_N)$ is shorthand for ${\sf Pr}(X_1=i_1,\ldots,X_N=i_N) $ and likewise ${\sf Pr}(i_N | i_1 \ldots, i_{N-1})$ for ${\sf Pr}(X_N=i_N | X_1=i_1 \ldots, X_{N-1}=i_{N-1})$.
In recommender systems, given a set of observed item ratings $X_1\ldots,X_{N-1}$ one can compute the conditional expectation of an unobserved rating given the observed ones
\[ {\mathbb E} ( X_N | i_1,\ldots,i_{N-1})  = \sum_{i_N = 1}^{I_N} i_N {\sf Pr}(i_N | i_1,\ldots,i_{N-1} ).\]

At this point the reader may wonder why we consider recovery from { three-dimensional} joint PMFs and not from {one- or two-dimensional} PMFs. It is well-known that recovery from { one-dimensional} marginal PMFs is  possible when all random variables are {known to be} independent. {In this case, the joint PMF is equal to the product of the individual one-dimensional marginals. Interestingly, recovery from one-dimensional marginals is also possible when the random variables are known to be fully dependent i.e., one is completely determined by the other. In this case, the joint PMF can be recovered if each one-dimensional marginal is a unique permutation of the other.} 

 {However, complete (in)dependence is unrealistic in statistical estimation and learning practice. In general it is not possible to recover a joint PMF from one-dimensional marginals.} An illustration for two variables is shown in Figure~\ref{fig:rank1}: ${\sf Pr}(i_1,i_2)$ can be represented as a matrix, and ${\sf Pr}(i_1)$,  ${\sf Pr}(i_2)$ are `projections' of the matrix along the row and column directions using the projector ${\bf 1}^T$ and ${\bf 1}$, respectively: {${\sf Pr}(i_1)=\sum_{i_2 = 1}^{I_2} {\sf Pr}(i_1,i_2)$ and  ${\sf Pr}(i_2)=\sum_{i_1 =1 }^{I_1} {\sf Pr}(i_1,i_2)$}.
In this case, if we denote ${\bf P}$ the matrix such that ${\bf P}(i_1,i_2)={\sf Pr}(X_1=i_1,X_2=i_2)$, then ${\rm rank}({\bf P})=r>1$ if $X_1$ and $X_2$ are not independent. From basic linear algebra, one can see that knowing ${\bf 1}^T{\bf P}$ and ${\bf P}{\bf 1}$ is not enough for recovering ${\bf P}$ in general -- since this is equivalent to solving a very underdetermined system of linear equations with $(I_1+I_2)\times r$ variables but only $I_1 + I_2$ equations. 
\begin{figure}[!t]
\centering
\includegraphics[width=0.4\linewidth]{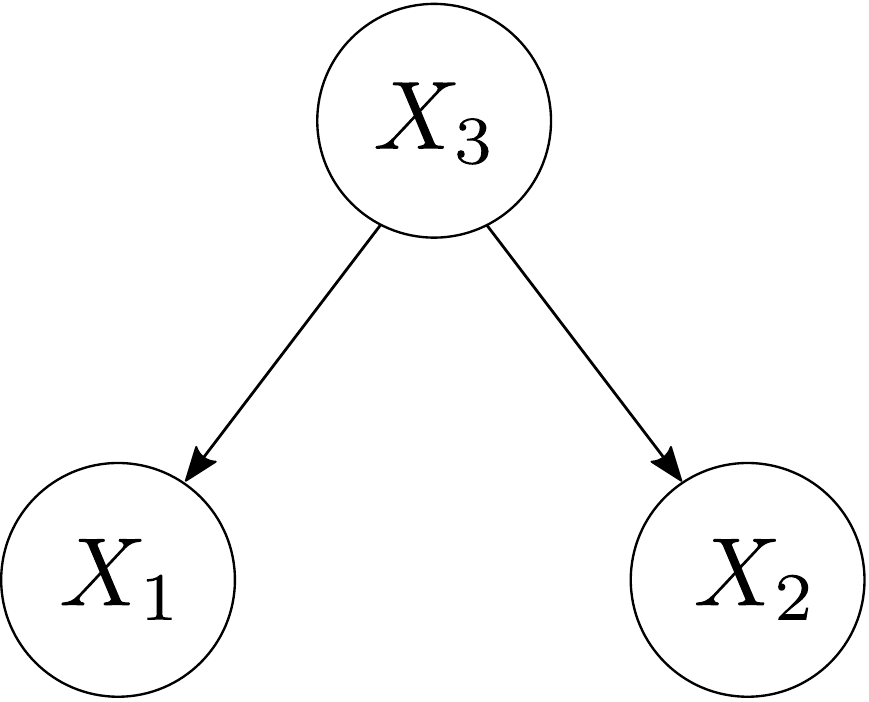}
\caption{Bayesian network of three variables.}
\label{fig:simple_bayes}
\end{figure}

What if we know {two-dimensional} marginals?  When the given random variables obey a probabilistic graphical model, and a genie reveals that model to us, then estimating a {high-dimensional} joint PMF from {two-dimensional} marginals may be possible. An example is shown in Figure~\ref{fig:simple_bayes}. If we know {\it a priori} that random variables $X_1$ and $X_2$ are conditionally independent given $X_3$, one can verify that knowledge of ${\sf Pr}(X_1=i_1,X_3=i_3)$ and ${\sf Pr}(X_2=i_2,X_3=i_3)$ is sufficient to recover ${\sf Pr}(X_1=i_1,X_2=i_2,X_3=i_3)$. However, this kind of approach hinges on knowing the probabilistic graph structure. Unfortunately, genies are hard to come by in real life, and learning the graph structure from data is itself a very challenging problem in statistical learning~\cite{KoFrie2009}. 

In our problem setup, we do not assume any {\it a priori} knowledge of the graph structure, and in this sense we have a `blind' joint PMF recovery problem. Interestingly, under certain conditions, this is no hindrance.

\section{Preliminaries}
\label{sec:cpd_pmf}
Our framework is heavily based on low-rank  tensor  factorization and its nice identifiability properties.
To facilitate our later discussion, we briefly introduce pertinent aspects of tensors in this section.
\subsection{Rank Decomposition}
An $N$-way tensor $\underline{\mathbf{X}} \in \mathbb{R}^{I_1 \times I_2 \times \cdots \times I_N}$ 
is a data array whose elements are indexed by $N$ indices. A {two-way} tensor is a matrix, whose elements have two indices; i.e., ${\bf X}(i,j)$ denotes the $(i,j)$-th element of the matrix ${\bf X}$. If a matrix ${\bf X}$ has rank $F$, it admits a rank decomposition ${\bf X}=\sum_{f=1}^F{\bf A}_1(:,f)\circ{\bf A}_2(:,f)= {\bf A}_1{\bf A}_2^T$
where we have ${\bf A}_n=[{\bf A}_n(:,1),\ldots,{\bf A}_n(:,F)]$ and $\circ$ denotes the outer product of two vectors, i.e.,
$[{\bf x}\circ {\bf y}](i,j)={\bf x}(i){\bf y}(j)$. Similarly, if an $N$-way tensor $\underline{\bf X}$ has rank $F$, it
admits the following rank decomposition:
\begin{equation}\label{eq:tensor_decomp}
\underline{\mathbf{X}} = \sum_{f=1}^F\mathbf{A}_1(:,f)\circ \mathbf{A}_2(:,f) \circ \cdots  \circ \mathbf{A}_N(:,f),
\end{equation}
where $\mathbf{A}_n \in \mathbb{R}^{I_n \times F}$ and $F$ is the smallest number for which such a decomposition exists. For convenience, we use the notation $\underline{\mathbf{X}} = [\![ \mathbf{A}_1,\ldots,\mathbf{A}_N ]\!]$ to denote the decomposition. 
The above rank decomposition is also called the Canonical Polyadic Decomposition (CPD) or Parallel Factor Analysis (PARAFAC) model of a tensor. 
It is critical to note that every tensor admits a {CPD}, and that the rank $F$ is not necessarily smaller than $I_1,\ldots,I_N$ -- the latter is in sharp contrast to the matrix case~\cite{SiDeFu2017}.
\begin{figure}
	\centering
	\includegraphics[width=1\linewidth]{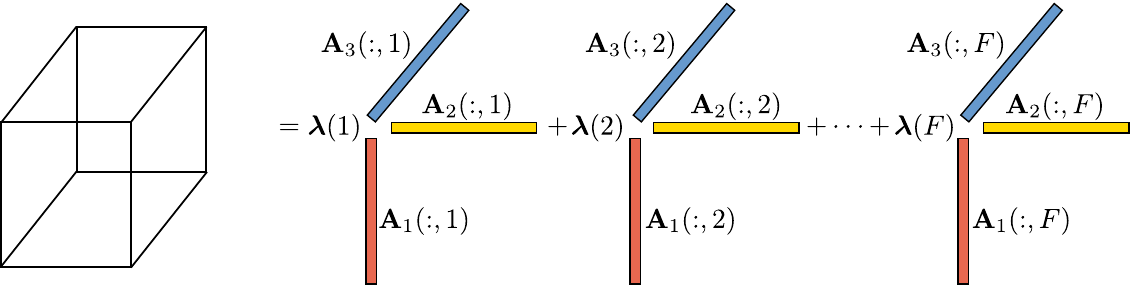}
	\caption{Illustration of the rank decomposition of a {three-way} tensor.}
	\label{fig:cpd}
\end{figure}
In the matrix case, it is easy to see that ${\bf X}(i_1,i_2)=\sum_{f=1}^F  {\bf A}_1(i_1,f)\mathbf{A}_2(i_2,f)$.
Similarly, for an $N$-way tensor we have $\underline{\mathbf{X}}(i_1,i_2,\ldots,i_N) = \sum_{f=1}^F  \prod_{n=1}^N \mathbf{A}_n(i_n,f)$. Sometimes one wishes to restrict the columns of ${\bf A}_n$'s to have unit norm (e.g., as in SVD). Therefore, the tensors can be represented as
\begin{equation}\label{eq:tensor_decomp_norm}
\underline{\mathbf{X}} = \sum_{f=1}^F{\bm \lambda}(f)\mathbf{A}_1(:,f)\circ \mathbf{A}_2(:,f) \circ \cdots  \circ \mathbf{A}_N(:,f),
\end{equation}
or, equivalently
\begin{equation}
\underline{\mathbf{X}}(i_1,i_2,\ldots,i_N) = \sum_{f=1}^F \boldsymbol{\lambda}(f)  \prod_{n=1}^N \mathbf{A}_n(i_n,f),
\label{eq:individ}
\end{equation}
where $\|{\bf A}_n(:,f)\|_p=1$ for {a certain} $p\geq 1$,  $\forall \;  n,f$, and ${\bm \lambda}=[{\bm \lambda}(1),\ldots,{\bm \lambda}(F)]^T$ with $\|{\bm \lambda}\|_0=F$ is employed to `absorb' the norms of columns. An illustration of a {three-way} tensor and its CPD is shown in Figure~\ref{fig:cpd}. Under such cases, we denote the $N$-way tensor as  $\underline{\mathbf{X}} = [\![ {\bm \lambda}, \mathbf{A}_1,\ldots,\mathbf{A}_N ]\!]$ -- again, in this expression, we have automatically assumed that $\|{\bf A}_n(:,f)\|_p=1$, $\forall \; n,f$ and a certain $p\geq1$. {We will refer to the decomposition of $\underline{\mathbf{X}}$ into nonnegative factors $\boldsymbol{\lambda} \in \mathbb{R}_+^F$, $\mathbf{A}_n \in \mathbb{R}_+^{I_n \times F}$ as nonnegative decomposition.}

%\blue{
%In many applications, we are interested in computing a rank decomposition of a nonnegative tensor $\underline{\mathbf{X}}$ using nonnegative factors $\mathbf{A}_n \in \mathbb{R}_+^{I_n \times F}$. The smallest number $F$ for which such a decomposition exists is called the nonnegative rank and is denoted as $ \textrm{rank}_+(\underline{\mathbf{X}})$. We refer to this decomposition as the nonnegative CPD of $\underline{\mathbf{X}}$. A
%nonnegative tensor may have different nonnegative and real ranks with $\textrm{rank}(\underline{\mathbf{X}}) \leq \textrm{rank}_+(\underline{\mathbf{X}})$}.

The following definitions will prove useful in the rest of the paper. We define the \emph{mode-$n$ matrix unfolding} of $\underline{\mathbf{X}}$ as the matrix ${\mathbf{X}}^{(n)}$ of size $\prod_{ \substack{k=1 \\ k\neq n}}^N I_k \times I_n$. We have that $\underline{\mathbf{X}}(i_1,i_2,\ldots,i_N) = {\mathbf{X}}^{(n)}(j,i_n)$, where
\begin{equation*}
j = 1 + \sum_{ \substack{k=1 \\ k \neq n} }^N (i_k-1) J_k \; \text{with} \; J_k = \prod_{  \substack{m=1 \\ m \neq n}}^{k-1} I_m.
\end{equation*}
In terms of the CPD factors, the mode-$n$ matrix unfolding can be expressed as
\begin{equation}
{\mathbf{X}}^{(n)} =  \left( \underset{j\neq n}{ \underset{j=1}{ \overset{N}{\odot}}} \mathbf{A}_j \right) \textrm{diag}(\boldsymbol{\lambda}) \mathbf{A}_n^T,
\end{equation}
where
$\underset{j\neq n}{ \underset{j=1}{ \overset{N}{\odot}}} \mathbf{A}_j = \mathbf{A}_N \odot \cdots \odot \mathbf{A}_{n+1} \odot \mathbf{A}_{n-1} \odot \cdots \odot \mathbf{A}_1.$

We can also express a tensor in a vectorized form $\underline{\mathbf{X}}(i_1,i_2,\ldots,i_N) = {\mathbf{x}}(j)$, where
\begin{equation*}
j = 1 + \sum_{k=1 }^N (i_k-1) J_k \; \text{with} \; J_k = \prod_{m=1}^{k-1} I_m.
\end{equation*}
In terms of the CPD factors, the vectorized form of a tensor can be expressed as
\begin{equation}
\text{vec}(\underline{\mathbf{X}})=  \left( { \underset{j=1}{ \overset{N}{\odot}}} \mathbf{A}_j \right) \boldsymbol{\lambda}.
\end{equation}

\subsection{Uniqueness of Rank Decomposition}
A distinctive feature of tensors is that they have \emph{essentially} unique CPD under mild conditions -- even when $F$ is much larger than $I_1,\ldots,I_N$.
To continue our discussion, let us first formally define what we mean by essential uniqueness of rank decomposition of tensors.
\begin{Def}
	(Essential uniqueness) For a tensor $\underline{\mathbf{X}}$ of (nonnegative) rank $F$, we say that a nonnegative decomposition $\underline{\mathbf{X}} = [\![{\bm \lambda}, \mathbf{A}_1, \ldots, \mathbf{A}_N ]\!]$, {$\boldsymbol{\lambda} \in \mathbb{R}_+^F$, $\mathbf{A}_n \in \mathbb{R}_+^{I_n \times F}$} is essentially unique if the factors are unique up to a common permutation. This means that if there exists another nonnegative decomposition $\underline{\mathbf{X}} = [\![ \widehat{\boldsymbol{\lambda}},\widehat{\mathbf{A}}_1, \ldots, \widehat{\mathbf{A}}_N ]\!]$, then, there exists a permutation matrix $\boldsymbol{\Pi}$ such that 
$
\widehat{\mathbf{A}}_n = \mathbf{A}_n \boldsymbol{\Pi}, \forall n\in[N] \; \text{and} \; \widehat{\boldsymbol{\lambda}} = \boldsymbol{\Pi}^T \boldsymbol{\lambda}.
$
\end{Def}
In other words, if a tensor has an essentially unique nonnegative CPD, then the only ambiguity is column permutation of the column-normalized factors $\{\mathbf{A}_n\}_{n=1}^N$, which simply amounts to a permutation of the rank-one `chicken feet' outer products (rank-one tensors) in Fig. \ref{fig:cpd}, that is clearly unavoidable\footnote{Generally, there is also column scaling / counter-scaling ambiguity \cite{SiDeFu2017}: a red column can be multiplied by $\gamma$ and the corresponding yellow column divided by $\gamma$ without any change in the outer product. There is no scaling ambiguity for {\em nonnegative} column-normalized representation $\underline{\mathbf{X}} = [\![{\bm \lambda},\mathbf{A}_1, \ldots, \mathbf{A}_N ]\!]$, where there is obviously no sign ambiguity and all scaling is `absorbed' in ${\bm \lambda}$.}.
Regarding the essential uniqueness of tensors, let us consider the {three-way} case first. The following is arguably the most well-known uniqueness condition that was revealed by Kruskal in 1977.

\begin{Lemma}\label{lem:kruskal}
\cite{Kru1977} Let $\underline{\mathbf{X}} = [\![{\bm \lambda}, \mathbf{A}_1,\mathbf{A}_2,\mathbf{A}_3 ]\!]$, where $\mathbf{A}_1 \in \mathbb{R}^{I_1 \times F}$, $\mathbf{A}_2 \in \mathbb{R}^{I_2 \times F}$, $\mathbf{A}_3 \in \mathbb{R}^{I_3 \times F}$. If $k_{\mathbf{A}_1} + k_{\mathbf{A}_2} + k_{\mathbf{A}_3} \geq 2F + 2$ then $\textrm{rank}(\underline{\mathbf{X}}) = F $ and the decomposition of $\underline{\mathbf{X}}$ is essentially unique. 
\end{Lemma}
Here, $k_{\mathbf{A}}$ denotes the Kruskal rank of the matrix $\mathbf{A}$ which is equal to the largest integer such that every subset of $k_{\mathbf{A}}$ columns are linearly independent. Lemma~\ref{lem:kruskal} implies the following generic result: 
The decomposition $\underline{\mathbf{X}} = [\![{\bm \lambda}, \mathbf{A}_1,\mathbf{A}_2,\mathbf{A}_3 ]\!]$ is essentially unique, almost surely, if
\begin{equation}\label{lem:thm1}
\textrm{min}(I_1,F) + \textrm{min}(I_2,F) + \textrm{min}(I_3,F) \geq 2F + 2.
\end{equation}
This is because ${k}_{{\bf A}_n}=\min(I_n,F)$ with probability one if the elements of ${\bf A}_n$ are generated following a certain absolutely continuous distribution. More relaxed and powerful uniqueness conditions have been proven in recent years.

\begin{Lemma}\label{lem:generic_1}\cite{ChiOtta2012}, ~\cite{DoDe2015} { Let $\underline{\mathbf{X}} = [\![\boldsymbol{\lambda} ,\mathbf{A}_1,\mathbf{A}_2,\mathbf{A}_3 ]\!]$ }, where $\mathbf{A}_1 \in \mathbb{R}^{I_1 \times F}$, $\mathbf{A}_2 \in \mathbb{R}^{I_2 \times F}$, $\mathbf{A}_3 \in \mathbb{R}^{I_3 \times F}$, $ I_1 \leq I_2 \leq I_3$, {$ I_1 \geq 3$} and $F \leq I_3$. Then, $\textrm{rank}(\underline{\mathbf{X}}) = F $ and the decomposition of $\underline{\mathbf{X}}$ is essentially unique, almost surely, if and only if $ F \leq (I_1 -1)(I_2 -1)$.
\end{Lemma}

\begin{Lemma}\label{lem:generic_2}
\cite{ChiOtta2012} Let $\underline{\mathbf{X}} = [\![{\bm \lambda}, \mathbf{A}_1,\mathbf{A}_2,\mathbf{A}_3 ]\!]$, where $\mathbf{A}_1 \in \mathbb{R}^{I_1 \times F}$, $\mathbf{A}_2 \in \mathbb{R}^{I_2 \times F}$, $\mathbf{A}_3 \in \mathbb{R}^{I_3 \times F}$, $ I_1 \leq I_2 \leq I_3$. Let $\alpha,\beta$ be the largest integers such that $2^\alpha \leq I_1$ and $2^\beta \leq I_2$. If $F \leq 2^{\alpha + \beta -2}$ then the decomposition of $\underline{\mathbf{X}}$ is essentially unique almost surely. The condition also implies that if $F\leq \frac{(I_1+1)(I_2+1)}{16}$, then $\underline{\mathbf{X}}$ has a unique decomposition almost surely.
\end{Lemma}

There are many more different uniqueness conditions for {CPD}. The take-home point here is that the CPD model is essentially generically unique even if $F$ is much larger than $I_1,I_2,I_3$ -- so long it is less than maximal possible rank.
For example, in Lemma \ref{lem:generic_2}, $F$ can be as large as ${\cal O}(I_1I_2)$ (but not equal to $I_1 I_2$), and the CPD model is still unique.

\begin{Remark}
{\rm
{ We should mention that the above identifiability results are derived for tensors under a noiseless setup\footnote{In this context, noise will typically come from insufficient sample averaging in empirical frequency estimation.}. In addition, although the results are stated for real factor matrices, they are very general and also cover nonnegative ${\bf A}_n$'s due to the fact that the nonnegative orthant has positive measure. It follows that if a tensor is generated using random nonnegative factor matrices then under the noiseless setup, a plain CPD can recover the true nonnegative factors. On the other hand, in practice, instead of considering {\it exact} tensor decomposition, often {\it low-rank tensor approximation} is of interest, because of limited sample size and other factors. The best low-rank tensor approximation might not even exist in this case; fortunately, adding structural constraints on the latent factors can mitigate this, see ~\cite{QiCoLim2016}. In this work, our interest lies in revealing the fundamental limits of joint PMF estimation. Therefore, our analysis will be leveraging exact decomposition results, e.g., Lemmas~\ref{lem:generic_1}-\ref{lem:generic_2}. However, since the formulated problem naturally involves nonnegative latent factors, our computational framework utilizes this structural prior knowledge to enhance performance in practice.}}
\end{Remark}

\section{Naive Bayes Model: A Rank-decomposition Perspective}\label{sec:connection}

\begin{figure}[!t]
\centering
\includegraphics[height= 0.15 \textwidth ]{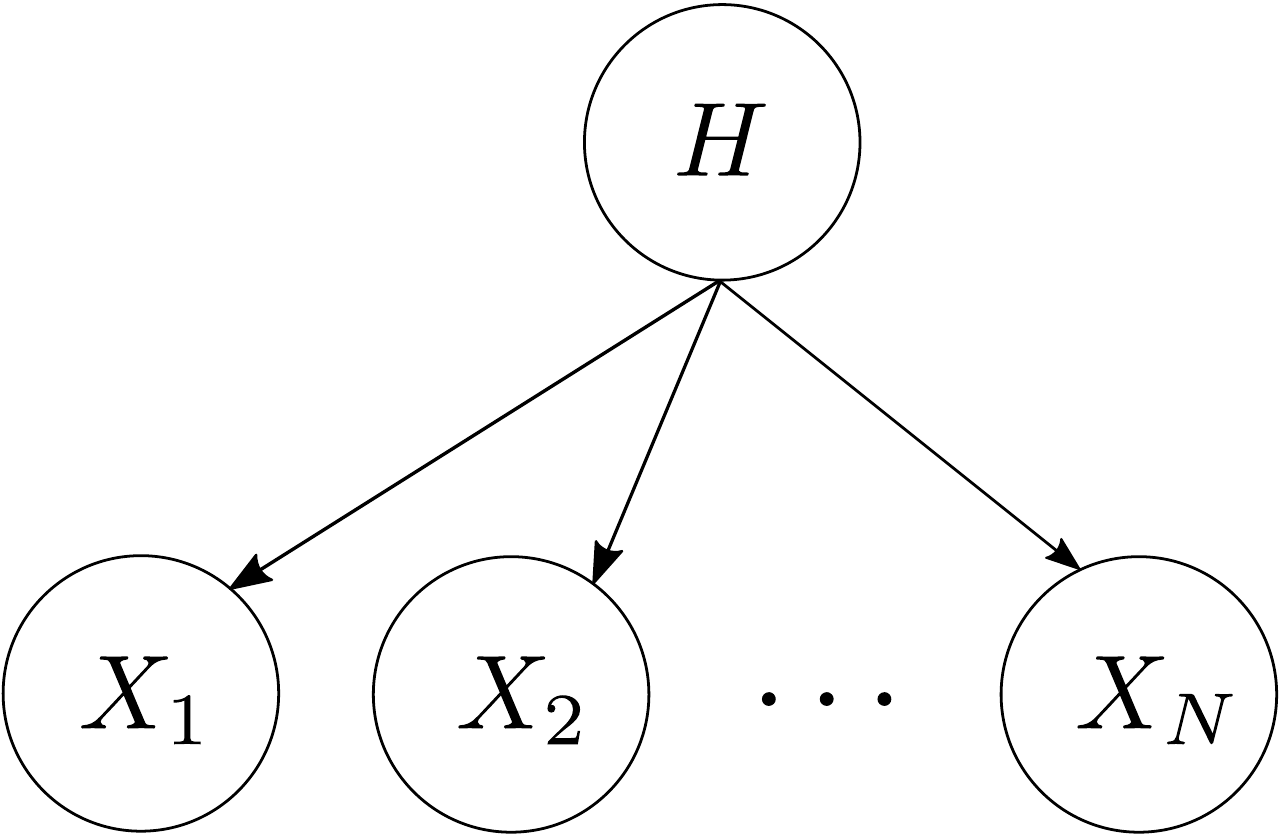}
\caption{Naive Bayes model.}
\label{fig:naive_model}
\end{figure}

We will show that {\em any} joint PMF admits a naive Bayes model {\em representation}, i.e., it can be generated from a latent variable model with just one hidden variable. The naive Bayes model postulates that there is a hidden discrete random variable $H$ taking $F$ possible values, such that given $H=h$ the discrete random variables $\{X_n\}_{n=1}^N$ are conditionally independent. It follows that the joint PMF of $\{X_n\}_{n=1}^N$ can be decomposed as
\begin{equation}\label{eq:pmf_latent_var}
\begin{aligned}
{\sf Pr}(i_1,i_2,\ldots,i_N) = \sum_{f=1}^F {\sf Pr}(f) \prod_{n=1}^N {\sf Pr}(i_n | f)
\end{aligned},
\end{equation}
where ${\sf Pr}(f) := {\sf Pr}(H = f) $ is the prior distribution of the latent variable $H$ and ${\sf Pr}(i_n | f) := {\sf Pr}( X_n = i_n | H = f) $ are the conditional distributions (Fig.~\ref{fig:naive_model}). 
The naive Bayes model in~\eqref{eq:pmf_latent_var} is also referred to as the latent class model~\cite{Zhang2004} and is the simplest form of a Bayesian network~\cite{KoFrie2009}. It has been employed in diverse applications such as classification~\cite{NgJo2002}, density estimation~\cite{LoDo2005} and crowdsourcing~\cite{DaSke1979}, just to name a few. 

An interesting observation is that the naive Bayes model can be interpreted as a special nonnegative polyadic decomposition. This was alluded to in \cite{ShaHa2005,LiCo2009} but not exploited for identifying the joint PMF from { lower-dimensional marginals}, as we do. Consider the element-wise representation in \eqref{eq:individ} and compare it with \eqref{eq:pmf_latent_var}: each column of the factor matrices can represent a conditional PMF and the vector $\boldsymbol{\lambda}$ contains the prior probabilities of the latent variable $H$, i.e., 
\begin{equation}\label{eq:link}
\mathbf{A}_n(i_n,f) = {\sf Pr}(i_n | f), \quad \boldsymbol{\lambda}(f) = {\sf Pr}(f).
\end{equation} 
This is a {\em special} nonnegative polyadic decomposition model because it restricts ${\bf 1}^T \boldsymbol{\lambda} = 1$. There is a subtle point however: the maximal rank $F$ in a CPD ({\em canonical} polyadic decomposition) model is bounded, but the number of latent states (latent alphabet size) for the naive Bayes model may exceed this bound. Even if the number of latent states is under the maximal rank bound, a naive Bayes model may be {\em reducible}, in the sense that there exists a naive Bayes model with fewer latent states that generates the same joint PMF. The net result is that {\em every} joint PMF admits a naive Bayes model {\em interpretation} with bounded $F$, and every naive Bayes model is or can be reduced to a  special CPD model. We have the following result.    

\begin{Prop}\label{prop:F_upperbound}
The maximum $F$ needed to represent an {\em arbitrary} PMF as a naive Bayes model is bounded by the following inequality
\begin{equation}\label{eq:F_up}
F\leq \underset{k}{\min}\left(\prod_{\substack{n=1 \\ n \neq k}}^N I_n\right).
\end{equation}
\end{Prop} 

{
\begin{IEEEproof}
Let $\underline{\mathbf{X}} \in \mathbb{R}_+^{I_1 \times I_2 \times I_3}$ denote a joint PMF of three random variables i.e., $\underline{\mathbf{X}}(i_1,i_2,i_3) = {\sf Pr}(X_1 = i_1, X_2 = i_2, X_3 = i_3)$. We define the following matrices
\begin{equation*}
\begin{aligned}
\mathbf{A}_1 &:= [\underline{\mathbf{X}}(:,:,1), \cdots, \underline{\mathbf{X}}(:,:,I_3)],  \\
\mathbf{A}_2 &:= [\mathbf{I}_{I_2 \times I_2}, \cdots, \mathbf{I}_{I_2 \times I_2}] = \textbf{1}_{I_3}^T \otimes \textbf{I}_{I_2 \times I_2},  \\
\mathbf{A}_3 &:= \textbf{I}_{I_3 \times I_3} \otimes \textbf{1}_{I_2}^T,
\end{aligned}
\end{equation*}
where $\mathbf{A}_1 \in \mathbb{R}_+^{I_1 \times I_2I_3}, \mathbf{A}_2 \in \mathbb{R}_+^{I_2 \times I_2I_3}, \mathbf{A}_3 \in \mathbb{R}_+^{I_3 \times I_2I_3}$ {and have used MATLAB notation $\underline{\mathbf{X}}(:,:,i_3)$ to denote the frontal slabs of the tensor $\underline{\mathbf{X}}$}. Additionally, $\mathbf{I}_{I_n \times I_n}$ denotes the identity matrix of size $I_n \times I_n$ and $\textbf{1}_{I_n}$ is a vector of all $1$'s of size $I_n$. Then every frontal slab of the tensor $\underline{\mathbf{X}}$ can be synthesized as $\underline{\mathbf{X}}(:,:,i_3) = \mathbf{A}_1 \textrm{diag}(\mathbf{A}_3(i_3,:)) \mathbf{A}_2^T.$
Upon normalizing the columns of matrix $\mathbf{A}_1$ such that they sum to one and absorbing the scaling in $\boldsymbol{\lambda}$, i.e., $\mathbf{A}_1=\mathbf{\widehat{A}}_1  \rm{diag}(\boldsymbol{\lambda}) $ we can decompose the tensor as $\underline{\mathbf{X}} = [\![\boldsymbol{\lambda},\mathbf{\widehat{A}}_1,\mathbf{A}_2,\mathbf{A}_3]\!]$. The number of columns of each factor is $I_2 I_3$. Due to role symmetry, by permuting the modes of the tensor it follows that we need at most $\textrm{min}(I_1I_2,I_2I_3,I_1I_3)$ columns for each factor for exact decomposition. 

The result is easily generalized to a {four-way} tensor $\underline{\mathbf{X}} \in \mathbb{R}_+^{I_1 \times I_2 \times I_3 \times I_4}$ by noticing that each slab  $\underline{\mathbf{X}}(:,:,:,i_4)$ is a {three-way} tensor and thus can be decomposed as $[\![\boldsymbol{\lambda}_{i_4},\widehat{\mathbf{A}}_{1,i_4},\mathbf{A}_{2,i_4},\mathbf{A}_{3,i_4}]\!]$ as before. We define
\begin{equation*}
\begin{aligned}
\boldsymbol{\lambda} &= [\boldsymbol{\lambda}_1^T,\cdots,\boldsymbol{\lambda}_{I_4}^T]^T, \\ 
\widehat{\mathbf{A}}_1 &= [\widehat{\mathbf{A}}_{1,1},  \cdots, \widehat{\mathbf{A}}_{1,I_4}], \quad
\mathbf{A}_2 = [\mathbf{A}_{2,1},  \cdots, \mathbf{A}_{2,I_4}], \\
\mathbf{A}_3 &= [\mathbf{A}_{3,1},  \cdots, \mathbf{A}_{3,I_4}], \quad
\mathbf{A}_4 = \mathbf{I}_{I_4} \otimes \mathbf{1}_{I_2I_3}^T.
\end{aligned}
\end{equation*}
The {four-way} tensor can therefore be decomposed as $[\![\boldsymbol{\lambda},\widehat{\mathbf{A}}_1,\mathbf{A}_2,\mathbf{A}_3,\mathbf{A}_4]\!]$. Due to symmetry, the number of columns of each factor is at most $\textrm{min}(I_1I_2I_3,I_2I_3I_4,I_1I_3I_4,I_1I_2I_4)$. By the same argument it follows that for a $N$-way tensor the bound on the nonnegative rank is $\underset{k}{\min}(\prod_{\substack{n=1 \\ n \neq k}}^N I_n)$.
\end{IEEEproof}}

\begin{shaded*}
The proof of Proposition~\ref{prop:F_upperbound} employs the same type of argument used to prove the upper bound on tensor rank. The main difference is in the normalization -- latent nonnegativity follows from data nonnegativity ``for free'' {since the latent factors used for constructing the CPD are either fibers drawn from the joint PMF itself, or from identity matrices or Kronecker products thereof}. While the proof is fairly straightforward for someone versed in tensor analysis, the implication of this proposition to probability theory is significant: it asserts that \emph{every} joint PMF can be represented by a naive Bayes model with a \emph{bounded} number of latent states $|{\cal H}|$. In fact, the connection between a naive Bayes model and CPD was utilized to approach some machine learning problems such as community detection and Gaussian Mixture Model (GMM) estimation in~\cite{AnGeHsu2014a}. However, in those cases, the hidden variable has a specific physical meaning (e.g., $H=f$ represents the $f$th community in community detection) and thus connection was established using a specific data generative model. Here, we emphasize that even when there is no physically meaningful $H$ or presumed generative model, {\em one can always represent an arbitrary joint PMF, possibly corresponding to a very complicated probabilistic graphical model, as a ``simple'' naive Bayes model with a bounded number of latent states} $F$. This result is very significant, also because it spells out that the latent structure of a probabilistic graphical model cannot be identified by simply assuming few hidden nodes; one has to limit the number of hidden node states as well.
\end{shaded*}

\begin{figure}[!t]
\centering
\includegraphics[height= 0.14 \textwidth]{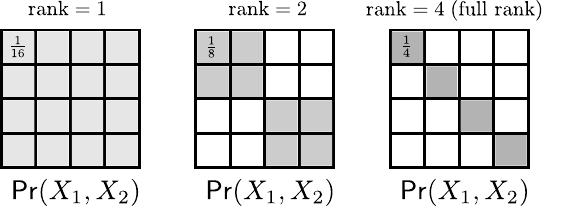}
\caption{Rank and independence.}
\label{fig:rank_ind}
\end{figure}

We should remark that although any joint PMF admits a naive Bayes representation, this does not mean that such representation is unique. Clearly, $F$ needs to be strictly smaller than the upper bound in \eqref{eq:F_up} to guarantee uniqueness (cf. Lemmas~\ref{lem:kruskal}-\ref{lem:generic_2}). Fortunately, many joint PMFs that we encounter in practice are relatively low-rank tensors, since random variables in the real world are only moderately dependent. This leads to an interesting connection between linear dependence/independence and statistical dependence/independence. To explain, let us consider the simplest case where $N=2$. In this case, we have
\begin{equation}
{\sf Pr}(i_1,i_2) = \sum_{f=1}^F {\sf Pr}(f)  {\sf Pr}(i_1 | f){\sf Pr}(i_2 | f).
\label{eq:pmf_latent_var_2}
\end{equation}
The {two-way model} corresponds to Nonnegative Matrix Factorization (NMF) and is related to Probabilistic Latent Semantic Indexing (PLSI)~\cite{Hofmann1999},~\cite{GaGo2005}. For the {two-way model}, independence of the variables implies that the probability matrix is rank-$1$. On the other hand, when the variables are fully dependent i.e., the value of one variable exactly determines the value of the other, the probability matrix is full-rank. However, low-rank does not necessarily mean that the variables are close to being independent as shown in Figure~\ref{fig:rank_ind}. There, a low rank probability matrix ($\text{rank}=2$) can also model highly dependent random variables. In practice, we expect that random variables will be neither independent nor fully dependent and we are interested in cases where the rank of the joint PMF is lower (and ideally much lower) than the upper bound given in Proposition \ref{prop:F_upperbound}.

As a sanity check, we conducted preliminary experiments on some real-life data. As anticipated, we verified that many joint PMFs are indeed low-rank tensors in practice. Table~\ref{table:real_data_rank} shows interesting results: The joint PMF of three movies over $5$ rating values was first estimated, using data from the MovieLens project. The joint PMF is then factored using a nonnegative CPD model with different rank values. One can see that with rank as low as $5$, the modeling error in terms of the relative error $\| \mathbf{X} - \widehat{\mathbf{X}} \|_F / \| \mathbf{X}\|_F$ is quite small, meaning that the low-rank modeling is fairly accurate. The same applies to two more datasets drawn from the UCI repository. 

\begin{table}[!t]
\begin{center}
\caption{Rel. error for different joint PMFs of $3$ variables.}\label{table:real_data_rank}
\begin{tabular}{l  c  c c }
&  \multicolumn{3}{c}{Rank ($F$)} \\
\hline
     & $5$  & $10$ & $15$ \\
\hline
INCOME     & $ 2.1 \times 10^{-2} $	& $5.5 \times 10^{-3}$ & $5.1 \times 10^{-3}$   \\
\hline
MUSHROOM   & $4.3 \times 10^{-2} $	& $2.4 \times 10^{-2}$ & $1.9 \times 10^{-2}$  \\
\hline
MOVIELENS   & $  1.8 \times 10^{-2}$	& $7.5 \times 10^{-3}$ & $4.1 \times 10^{-3}$  \\
\hline
\end{tabular}
\end{center}
\end{table}

\section{Joint PMF Recovery}
\subsection{General Procedures}

The key observation that enables our approach is that the marginal distribution of any subset of random variables is also a nonnegative CPD model. This is a direct consequence of the law of total probability. Marginalizing with respect to the $k$-th random variable we have that
\begin{align}
 \sum_{i_k=1}^{I_k}  {\sf Pr}(i_1, &   \ldots,i_N)  =  \sum_{f=1}^F \sum_{i_k=1}^{I_k} {\sf Pr}(f) \prod_{n=1}^N {\sf Pr}(i_n | f)  \nonumber 
\\ &   = \sum_{f=1}^F {\sf Pr}(f) \prod_{\substack{n=1 \\ n \neq k} }^N {\sf Pr}(i_n | f) \sum_{i_k=1}^{I_k} {\sf Pr}(i_k | f)    \nonumber \\  &  = \sum_{f=1}^F  {\sf Pr}(f) \prod_{\substack{n=1 \\ n \neq k} }^N {\sf Pr}(i_n | f), 
\end{align}
since $\sum_{i_n=1}^{I_n}  {\sf Pr}(i_n | f)  = 1$.

Consider the model in~\eqref{eq:pmf_latent_var} and assume that the marginal distributions  ${\sf Pr}(X_j=i_j,X_k=i_k,X_l=i_l) ,$ denoted ${\sf Pr}(i_j,i_k,i_l)$ for brevity, $\forall j,k,l \in [N],\; l>k>j$  are available and perfectly known. Then, there exists an exact decomposition of the form
\begin{equation}
{\sf Pr}(i_j,i_k,i_l) = \sum_{f=1}^{F}   {\sf Pr}(f) {\sf Pr}(i_j | f) {\sf Pr}(i_k | f) {\sf Pr}(i_l | f).
\end{equation}
The marginal distributions of triples of random variables satisfy $\underline{\mathbf{X}}_{jkl} = [\![ \boldsymbol{\lambda},\mathbf{A}_j,\mathbf{A}_k,\mathbf{A}_l ]\!]$, 
where $\{\mathbf{A}_n\}_{n=1}^N$ and ${\bm \lambda}$ are defined as in~\eqref{eq:link} and they satisfy $\mathbf{A}_l \geq \mathbf{0}$, $\mathbf{A}_k \geq \mathbf{0}$, $\mathbf{A}_j \geq \mathbf{0}$, $\mathbf{1}^T \mathbf{A}_l = \mathbf{1}^T$, $\mathbf{1}^T \mathbf{A}_k = \mathbf{1}^T$, $\mathbf{1}^T \mathbf{A}_j = \mathbf{1}^T$, $\boldsymbol{\lambda} > \mathbf{0} $, $\mathbf{1}^T \boldsymbol{\lambda} = 1$. 
Based on the connection between the naive Bayes model of { lower-dimensional} marginals and the joint PMF, we propose the following steps to recover the complete joint PMF from {three-dimensional} marginals.

\smallskip
\noindent
\fbox{
	\begin{minipage}{.95\linewidth}
		\textsf{\!\!\! \bf Procedure: Joint PMF Recovery From Triples}\\
		\; {\bf [S1]} Estimate $\underline{\bf X}_{jk\ell}$ from data;\\
		%\end{align*}}
		\; \!\!{\bf [S2]} Jointly factor $\underline{\mathbf{X}}_{jkl} = [\![ \boldsymbol{\lambda},\mathbf{A}_j,\mathbf{A}_k,\mathbf{A}_l ]\!]$ to estimate $\boldsymbol{\lambda},\mathbf{A}_j,\mathbf{A}_k,\mathbf{A}_l \; \forall$ $j,k,l$ using a CPD model with rank $F$;\\
		\; \!\!{\bf [S3]} Synthesize the joint PMF $\underline{\bf X}$ via ${\sf Pr}(i_1,i_2,\ldots,i_N) = \sum_{f=1}^F {\sf Pr}(f) \prod_{n=1}^N {\sf Pr}(i_n | f)$, w/  ${\sf Pr}(i_n | f)=\mathbf{A}_{n}(i_n,f)$, ${\sf Pr}(f) = \boldsymbol{\lambda}(f)$.
	\end{minipage}
}

\smallskip

One can see from step [S2], that if the individual factorization of at least one $\underline{\bf X}_{jkl}$ is unique, then the joint PMF is readily identifiable via [S3]. This is already very interesting. However, as we will show in Sec.~\ref{sec:ident}, we may identify the joint PMF even when the marginal tensors do not have unique {CPD}. The reason is that many marginal tensors share factors and we can exploit this to come up with much stronger identifiability results.

\subsection{Algorithm: Coupled Matrix/Tensor Factorization}
\label{sec:approach}
Before we discuss theoretical results such as identifiability of the joint PMF using {three or higher-dimensional} marginals,
we first propose an implementation of [S2] in the proposed procedure.
For brevity, we assume we have estimates of {three-dimensional} marginal distributions, i.e., we are given empirical estimates $\widehat{{\sf Pr}}(X_j=i_j,X_k=i_k,X_l=i_l), \; \forall j,k,l \in [N], \; l>k>j$, which we put in a tensor $\underline{\mathbf{X}}_{jkl}$ i.e., $\underline{\mathbf{X}}_{jkl}(i_j,i_k,i_l) = \widehat{{\sf Pr}}(X_j = i_j ,X_k = i_k ,X_l = i_l)$.

The method can be easily generalized to any type of { low-dimensional} marginal distributions. Under the assumption of a low-rank CPD model, every empirical marginal distribution of three random variables can be approximated as follows
\begin{equation}
\widehat{{\sf Pr}}(i_j,i_k,i_l) \approx \sum_{f=1}^{F}   {\sf Pr}(f) {\sf Pr}(i_j | f) {\sf Pr}(i_k | f) {\sf Pr}(i_l | f).
\end{equation}
Therefore, in order to compute an estimate of the full joint PMF, we propose solving the following optimization problem
\begin{equation} 
\begin{aligned}
&  \min_{ \{ \mathbf{A}_n\}_{n=1}^N,\boldsymbol{\lambda} } && \ \sum_{j} \sum_{k>j}\sum_{l>k} \frac{1}{2} \left \| \underline{\mathbf{X}}_{jkl} - [\![ \boldsymbol{\lambda},\mathbf{A}_j,\mathbf{A}_k,\mathbf{A}_l ]\!] \right \|_F^2 \\ 
& \text{subject to} 
& & \boldsymbol{\lambda}\geq \mathbf{0},~{\mathbf{1}}^T\boldsymbol{\lambda}=1, \\ 
&&& \mathbf{A}_n \geq \mathbf{0}, \; n=1,\ldots, N, \\ 
&&& {\mathbf{1}}^T \mathbf{A}_n = \mathbf{1}^T, \; n=1,\ldots,N.
\label{eq:coupled_tensor}
\end{aligned}
\end{equation}
The optimization problem in~\eqref{eq:coupled_tensor} is an instance of coupled tensor factorization. Coupled  tensor/matrix factorization is usually used as a way of combining various datasets that share dimensions and corresponding factor matrices~\cite{BeTaKu2014,AcKoDu2011}. Notice that in the case where we have estimates of {two-dimensional} marginals, the optimization problem in~\eqref{eq:coupled_tensor} corresponds to coupled matrix factorization. 
The optimization problem {\it per se} is very challenging and deserves developing sophisticated algorithms for handling it:
first, when the number of random variables ($N$) gets large, there is a large number of optimization variables (i.e., $\{{\bf A}_n\}_{n=1}^N$) to be determined in \eqref{eq:coupled_tensor} -- and each ${\bf A}_n$ is an $I_n\times F$ matrix where $I_n$ (the alphabet size of the $n$-th random variable) can be large. In addition, the probability simplex constraints impose some extra computational burden. Nevertheless, we found that, by carefully re-arranging terms, the formulated problem can be recast in convenient form and handled in a way that is reminiscent of the classical alternating least squares algorithm with constraints.
\begin{algorithm}[t]
\caption{Coupled Tensor Factorization Approach}
\begin{algorithmic}[1]
\STATEx \hspace*{-.4cm}
 \textbf{Input}: A discrete valued dataset $\mathbf{D}\in \mathbb{R}^{M \times N}$
\STATEx \hspace*{-.4cm}
\textbf{Output}: Estimates of $\{\mathbf{A}_n\}_{n=1}^N$ and $\boldsymbol{\lambda}$
\STATE Estimate $\underline{\mathbf{X}}_{j,k,l} \quad \forall j,k,l \in [N],\; l>k>j$ from data.
\STATE Initialize $\{\mathbf{A}_n\}_{n=1}^N$ and $\boldsymbol{\lambda}$ such that the probability simplex constraints are satisfied.
\REPEAT  
 \FORALL{$n \in [N]$} 
 \STATE{Solve optimization problem~\eqref{opt:coupled_tensor_sub1}} 
 \ENDFOR
 \STATE{Solve optimization problem~\eqref{opt:coupled_tensor_sub2}} 
\UNTIL{convergence criterion satisfied}
\end{algorithmic}
\label{alg:coupled_tensor_approach}
\end{algorithm}

The idea is that we cyclically update variables $\{\mathbf{A}_n\}_{n=1}^N$ and $\boldsymbol{\lambda}$ while fixing the remaining variables at their last updated values. Assume that we fix estimates $\boldsymbol{\lambda},\mathbf{A}_n$, $ \forall n \in [N] \setminus \{j\}$. Then, the optimization problem with respect to $\mathbf{A}_j$ becomes
\begin{equation}
\begin{aligned}
&  \min_{ \mathbf{A}_j} \sum_{k \neq j} \sum_{\substack{l\neq j \\ l > k}} && \frac{1}{2} \left \| \underline{\mathbf{X}}_{jkl} - [\![ \boldsymbol{\lambda},\mathbf{A}_j,\mathbf{A}_k,\mathbf{A}_l ]\!] \right \|_F^2 \\
& \text{subject to} & & \mathbf{A}_j \geq \mathbf{0},~{\mathbf{1}}^T \mathbf{A}_j = \mathbf{1}^T.   
\end{aligned}
\end{equation}
Note that we have dropped the terms that do not depend on $\mathbf{A}_j$. By using the \emph{mode-$1$ matrix unfolding} of each tensor $\underline{\mathbf{X}}_{jkl}$, the problem can be equivalently written as
\begin{equation}
\begin{aligned}
&  \min_{ \mathbf{A}_j} \sum_{k \neq j} \sum_{\substack{l\neq j \\ l > k}}  && \frac{1}{2} \left \| \mathbf{X}_{jkl}^{(1)} - (\mathbf{A}_l \odot \mathbf{A}_k) \textrm{diag}(\boldsymbol{\lambda}) \mathbf{A}_j^T \right \|_F^2 \\
& \text{subject to} & & \mathbf{A}_j \geq \mathbf{0}, ~ {\mathbf{1}}^T \mathbf{A}_j = \mathbf{1}^T, 
\end{aligned}
\label{opt:coupled_tensor_sub1}
\end{equation}
which is a least-squares problem with respect to matrix $\mathbf{A}_j$ under probability simplex constraints on its columns. The optimization problem has the same form for each factor $\mathbf{A}_n$ due to role symmetry. In order to update $\boldsymbol{\lambda}$ we solve the following optimization problem
\begin{equation}
\begin{aligned}
&  \min_{ \boldsymbol{\lambda} } \sum_{j} \sum_{k>j} \sum_{l>k}  && \frac{1}{2} \left \| \text{vec}(\underline{\mathbf{X}}_{jkl}) - (\mathbf{A}_l \odot \mathbf{A}_k \odot \mathbf{A}_j) \boldsymbol{\lambda}  \right \|_2^2 \\
& \text{subject to} & & \boldsymbol{\lambda} \geq \mathbf{0},~ {\mathbf{1}}^T \boldsymbol{\lambda} = 1.  \\
\end{aligned}
\label{opt:coupled_tensor_sub2}
\end{equation}

Both Problems~\eqref{opt:coupled_tensor_sub1} and \eqref{opt:coupled_tensor_sub2} are linearly constrained quadratic programs, and can be solved to optimality by many standard solvers. Here, we propose to employ the {\it Alternating Direction Method of Multipliers} (ADMM) to solve these two sub-problems because of its flexibility and effectiveness in handling large-scale tensor decomposition~\cite{HuSiLi2016,FuHuMa2015}. Details of the ADMM algorithm for solving Problems~\eqref{opt:coupled_tensor_sub1}-\eqref{opt:coupled_tensor_sub2} can be found in the Appendix~\ref{ap:algorithm}.
The whole procedure is listed in Algorithm~\ref{alg:coupled_tensor_approach}. As mentioned, the algorithm is easily modified to cover the cases where { higher-dimensional} marginals or pairwise marginals are given, and thus these cases are omitted.

\section{{ Joint PMF} Identifiability Analysis}\label{sec:ident}

In this section, we study the conditions under which we can identify ${\sf Pr}(i_1,\ldots,i_N)$ from marginalized {lower-dimensional} distributions. For brevity, we focus on  {three-dimensional as lower-dimensional} distributions, and even though many more results are possible, we concentrate here on the case $I_n=I \; \forall n \in[N]$ for ease of exposition and manuscript length considerations. Similar type of analysis applies when $I_1,\ldots,I_N$ are different, however the analysis should be customized to properly address particular cases. Our aim here is to convey the spirit of what is possible in terms of identifiability results, as we cannot provide an exhaustive treatment (there are combinatorially many cases, clearly). 

Obviously, if $\underline{\bf X}_{jkl}$ is individually identifiable for each combination of $j,k,l$, then, ${\sf Pr}(i_j|f)$, ${\sf Pr}(i_k|f)$, ${\sf Pr}(i_l|f)$, and ${\sf Pr}(f)$ are identifiable. This means that given  {three-dimensional marginal distributions}, ${\sf Pr}(i_1,\ldots,i_N)$ is generically identifiable if $F \leq \frac{3I-2}{2}$ assuming that $I_n = I \; \forall n \in[N]$. This can be readily shown by invoking Lemma~\ref{lem:kruskal}, equation ~\eqref{lem:thm1}, and the link between the naive Bayes model and tensor factorization discussed in Sec.~\ref{sec:connection}. Note that $F \leq \frac{3I-2}{2}$ is already not a bad condition, since in many cases we have approximately low-rank tensors in practice. However, since we have many factor-coupled $\underline{\bf X}_{jkl}$'s, this identifiability condition can be significantly improved. We have the following theorems.

{
\begin{Theorem}\label{thm:proposed_1}
	Assume that ${\sf Pr}( i_n| f), \; \forall n\in [N]$ are drawn from an absolutely continuous distribution, that $I_1=\ldots=I_N=I$, and that the joint PMF ${\sf Pr}(i_1,\ldots,i_N)$ can be represented using a naive Bayes model of rank $F$. If $N \leq I$ then, ${\sf Pr}(i_1,\ldots,i_N)$ is almost surely (a.s) identifiable from the ${\sf Pr}(i_j,i_k,i_l)$'s if
		\[F \leq I(N-2) \]
If $N>I$ then, ${\sf Pr}(i_1,\ldots,i_N)$ is  a.s. identifiable from the ${\sf Pr}(i_j,i_k,i_l)$'s if
\[F \leq \left( \lfloor \frac{\sqrt{NI-1}}{I}\rfloor I -1 \right )^2 \]
\end{Theorem}
\begin{IEEEproof}
	The proof is relegated to Appendix~\ref{ap:ident_results}.
\end{IEEEproof}
}

{
\begin{Theorem}\label{thm:proposed_2}
	Assume that ${\sf Pr}(i_n| f), \; \forall n\in [N]$ are drawn from an absolutely continuous distribution, that $I_1=\ldots=I_N=I$, and that the joint PMF ${\sf Pr}(i_1,\ldots,i_N)$ can be represented using a naive Bayes model of rank $F$.  {Let $\alpha$ be the largest integer such that $2^\alpha\leq \lfloor{\frac{N}{3}}\rfloor I$. Then, ${\sf Pr}(i_1,\ldots,i_N)$ is a.s. identifiable from the ${\sf Pr}(i_j,i_k,i_l)$'s if
	\[F\leq  4^{\alpha -1} \] which is implied by
	\[F\leq  \frac{(\lfloor{\frac{N}{3}}\rfloor I+1)^2}{16}.  \]
}\end{Theorem}

\begin{IEEEproof}
	The proof is relegated to Appendix~\ref{ap:ident_results}.
\end{IEEEproof}
}
The rank bounds in Theorems~\ref{thm:proposed_1}-\ref{thm:proposed_2} are nontrivial, albeit far from the maximal attainable rank for the cases considered.  Recalling that higher-order tensors  are identifiable for higher ranks, a natural question is whether knowledge of  {four- or higher-dimensional} marginals can further enhance identifiability of the complete joint PMF.  The next theorem shows that the answer is affirmative.

\begin{Theorem}
		Assume that  ${\sf Pr}(i_n| f), \; \forall n\in [N]$ are drawn from an absolutely continuous distribution, that $I_1=\ldots=I_N=I$, and that the joint PMF ${\sf Pr}(i_1,\ldots,i_N)$ can be represented using a naive Bayes model of rank $F$.  Further assume that $\mathcal{S} = [N]$ can be partitioned into $4$ disjoint subsets denoted by ${\cal S}_1,\ldots,{\cal S}_4$ such that the  {four-dimensional marginals} ${\sf Pr}(i_j,i_k,i_l,i_m), \; \forall j \in \mathcal{S}_1, \forall k \in \mathcal{S}_2,\forall l \in \mathcal{S}_3,\forall m \in \mathcal{S}_4$ are available. Then, the joint PMF ${\sf Pr}(i_1,\ldots,i_N)$ is a.s. identifiable if
	\begin{equation*}
	\begin{aligned}
		&F \leq I^2 |\mathcal{S}_3||\mathcal{S}_4|,\\
		&2F(F-1) \leq I^2 |\mathcal{S}_1||\mathcal{S}_2|(I|\mathcal{S}_1| - 1)(I|\mathcal{S}_2|-1).
	\end{aligned}
	\end{equation*}
	\label{thm:proposed_3}
\end{Theorem}
\begin{IEEEproof}
	The proof is relegated to Appendix~\ref{ap:ident_results}.
\end{IEEEproof}

The conditions of Theorem~\ref{thm:proposed_3} are satisfied for much higher rank than those of Theorems~\ref{thm:proposed_1}-\ref{thm:proposed_2} as shown in Tables~\ref{table:Rank_Bound1}-\ref{table:Rank_Bound2}. The results related to the  {four-dimensional} marginals are obtained following Theorem~\ref{thm:proposed_3} via checking all possible partitions. The caveat is that one may need many more samples to reliably estimate the  {four-dimensional} marginals. Nevertheless, the theorems that we present in this section offer insights regarding the choice of  {lower-dimensional} marginals to work with -- such choice depends on the size of the alphabet of each variable ($I$) and the number of variables ($N$) as well as the amount of available data samples.

\begin{table}[!t]
\begin{center}
\caption{Rank bounds for generic identifiability  $(I=3)$.}
\label{table:Rank_Bound1}
\begin{tabular}{c  c  c  c c c}
&  \multicolumn{5}{c}{Number of Variables ($N$)}      \\
\hline
      &  $6$ & $10$	& $20$ & $40$ & $80$   \\
\hline
Triples     &  $4$ & $ 7 $	& $27$ & $105$ & $ 410$   \\
\hline
Quadruples  &  $10$ & $36$	& $179$ & $729$ & $2916$  \\
\hline
\end{tabular}
\end{center}
\end{table}

\begin{table}[!t]
\begin{center}
\caption{Rank bounds for generic identifiability $(N=6)$.}
\label{table:Rank_Bound2}
\begin{tabular}{c  c  c  c c c}
&  \multicolumn{5}{c}{Alphabet size ($I$)}      \\
\hline
       		&   $6$ & $10$& $20$ & $40$  & $80$ \\
\hline
Triples     &   $24$ & $40$	& $105$  & $410$ & $1620$  \\
\hline
Quadruples  &   $45$ & $131$	& $544$  & $2220$  & $8966$  \\
\hline
\end{tabular}
\end{center}
\end{table}
\begin{Remark}
\rm
 {
The above results rely on Lemmas~\ref{lem:generic_1},~\ref{lem:generic_2} and concern the identifiability of a generic choice of parameters; i.e., the parameters are assumed to be drawn randomly from a jointly continuous distribution. At this point one may wonder whether this is a realistic assumption in practice. For example, in some latent model identification problems a hidden variable has specific physical meaning and an observed variable may not depend on the state of the hidden variable for one or more of its values. Consider a Hidden Markov Model (HMM) where we denote the observed variable at time $t$ as ${X}_t$ and the hidden state is ${S}_t$. The conditional distribution ${\mathbf{\widetilde{A}}_t(i,s) := {\sf Pr}(X_t = i | S_t = s)}$ may be the same for two different values $s_1$ and $s_2$ of the hidden state $S_t$. In such a case, the Kruskal rank of matrix $\mathbf{\widetilde{A}}_t$ would be equal to $1$, thereby rendering the deterministic identifiability condition (Lemma~\ref{lem:kruskal}) useless. Do note, however, that in our setting the latent variable $H$ does not necessarily have a physical interpretation; the CPD is just a convenient `universal' parametrization of the joint PMF. Therefore the conditional distribution of an observed variable may be the same for two values of the hidden state, but it may still depend on the value of the `virtual' global latent variable $H$, and hence recovery of the the joint PMF using lower-dimensional marginals could still be possible.}
\end{Remark}

\section{Numerical Results}
\label{sec:results}

\begin{figure*}[!t]
\centering
\includegraphics[height= 0.24 \textwidth]{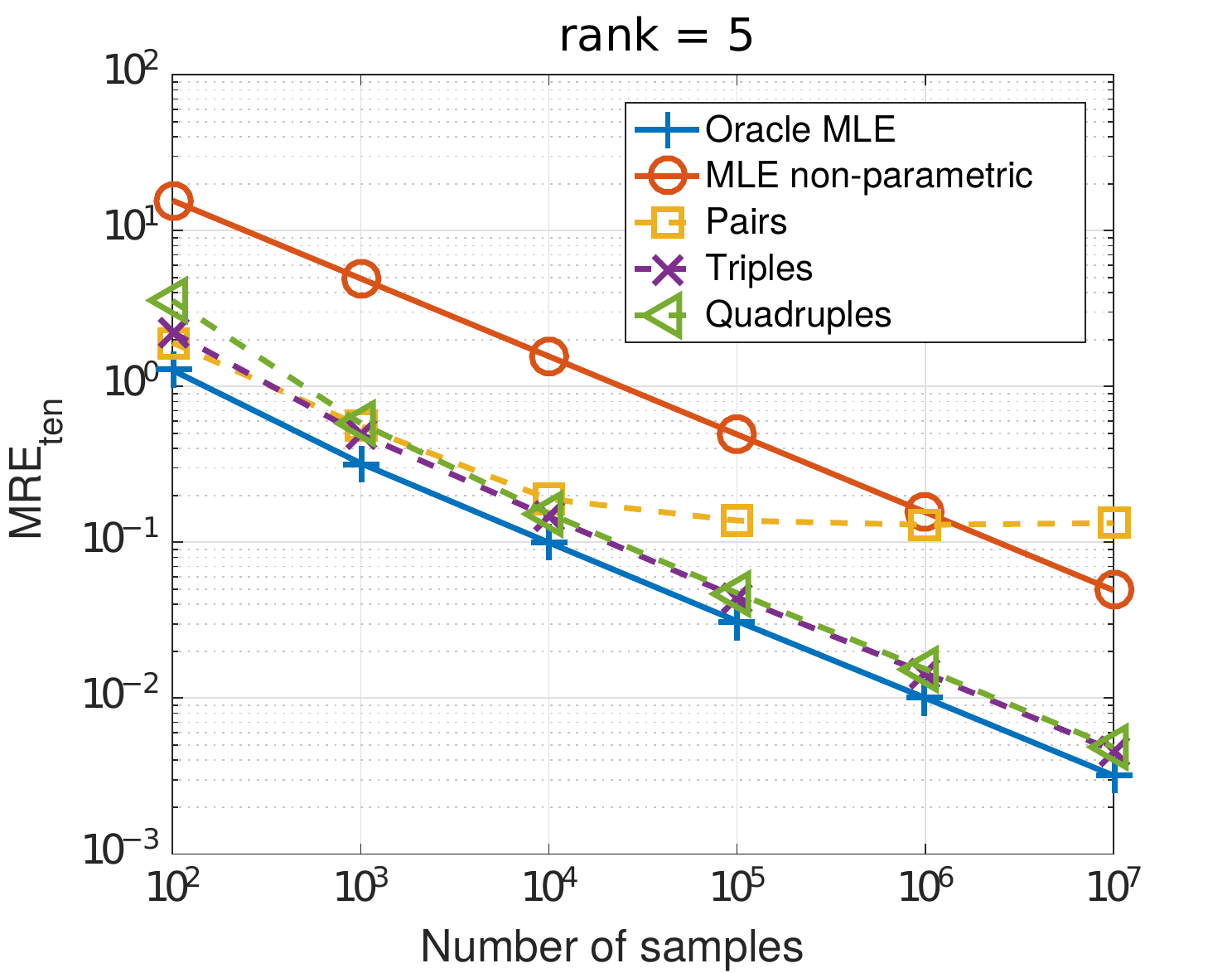}
\includegraphics[height= 0.24 \textwidth]{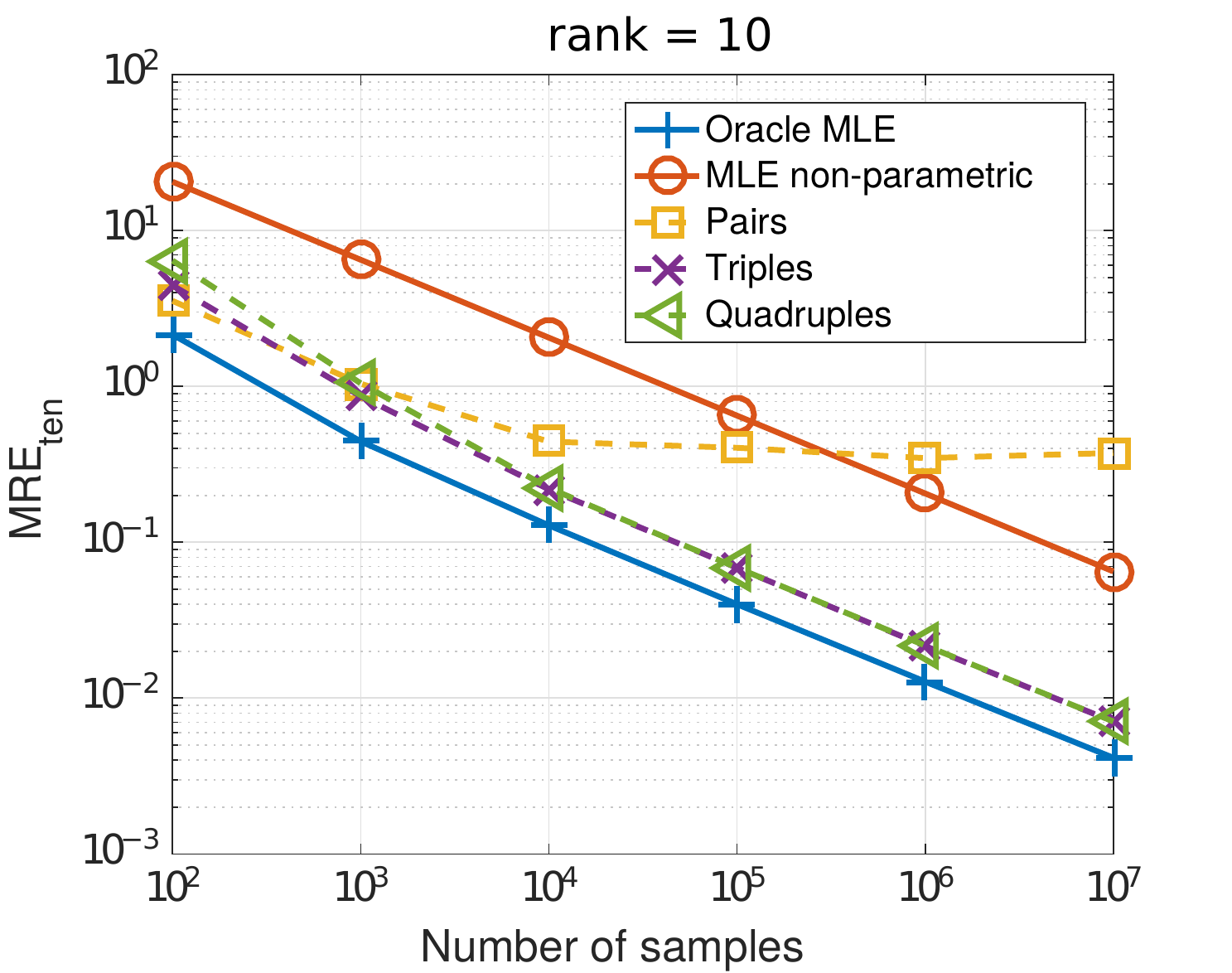}
\includegraphics[height= 0.24 \textwidth]{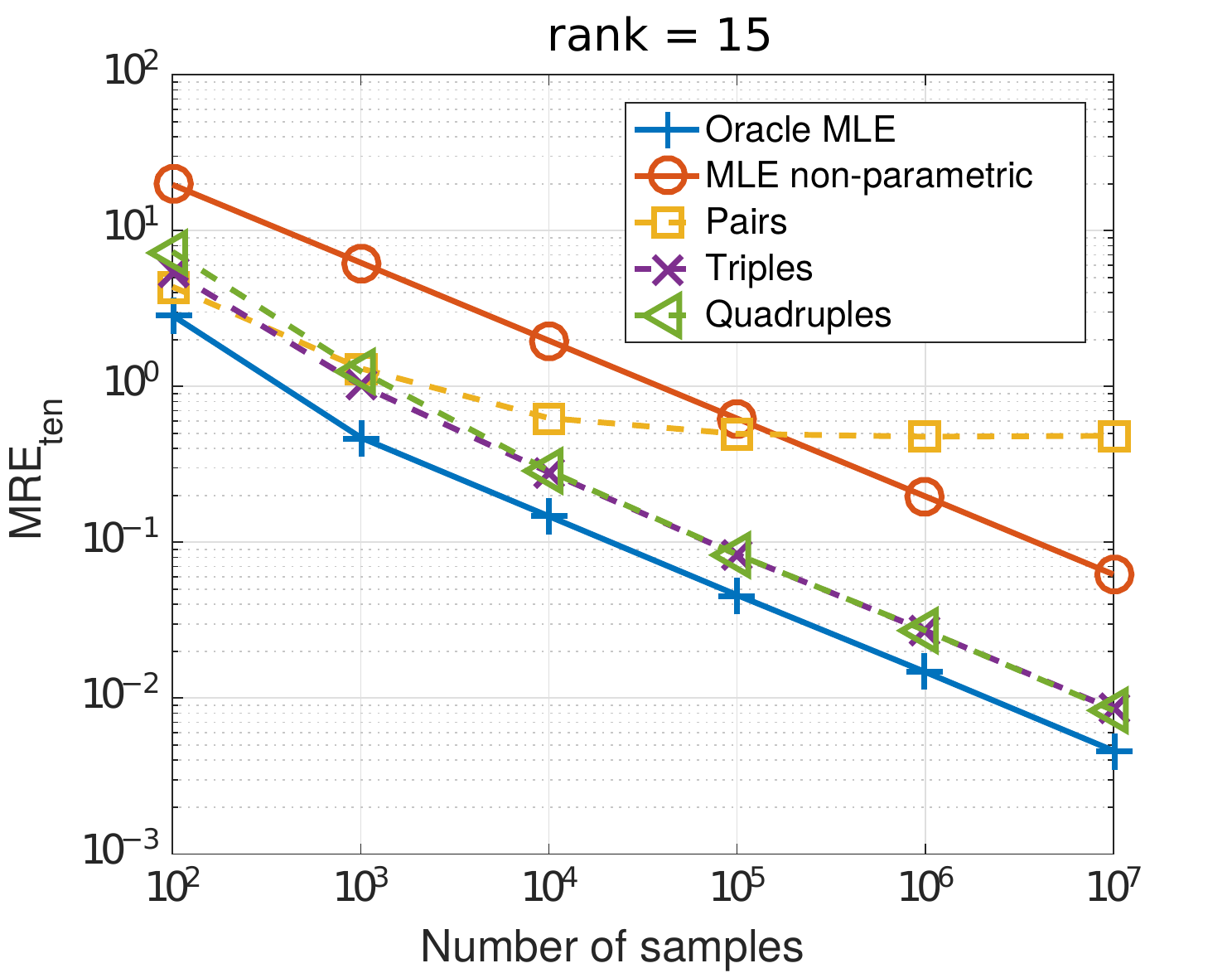}
\caption{Mean relative error of the estimated joint PMF under different number of available samples.}
\label{fig:synthetic_data}
\end{figure*}
\begin{figure*}[!t]
\centering
\includegraphics[height= 0.24 \textwidth]{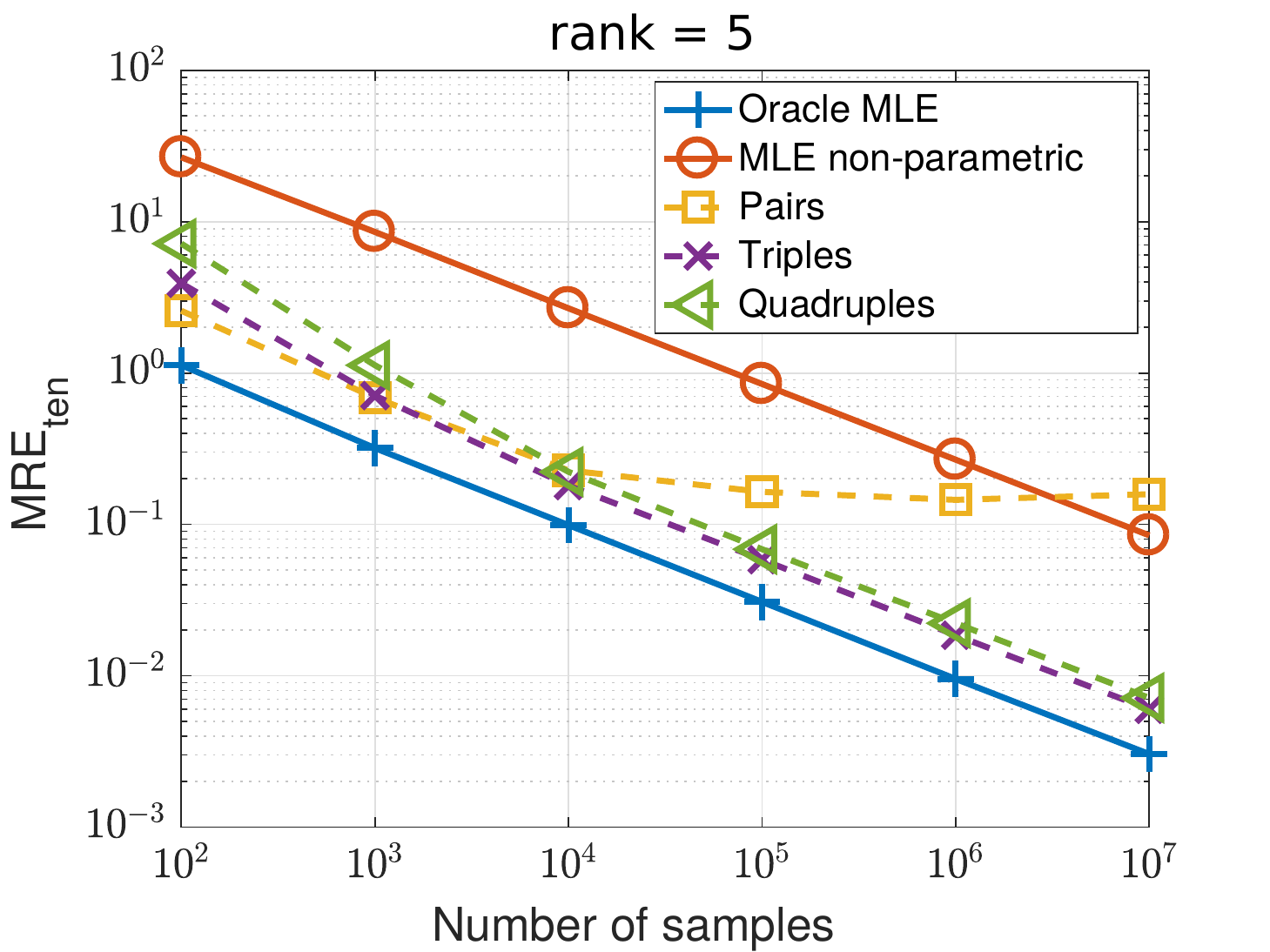}
\includegraphics[height= 0.24 \textwidth]{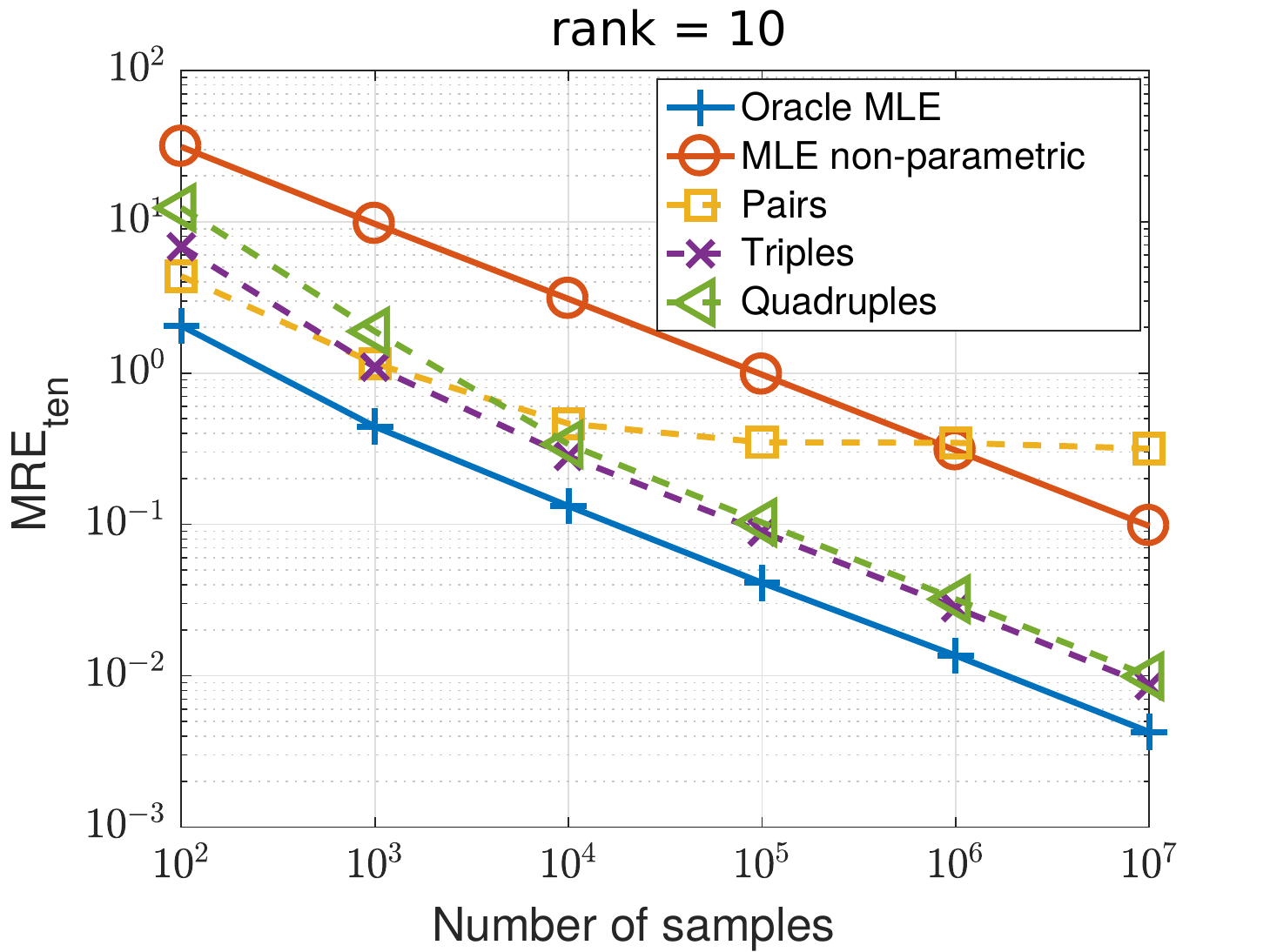}
\includegraphics[height= 0.24 \textwidth]{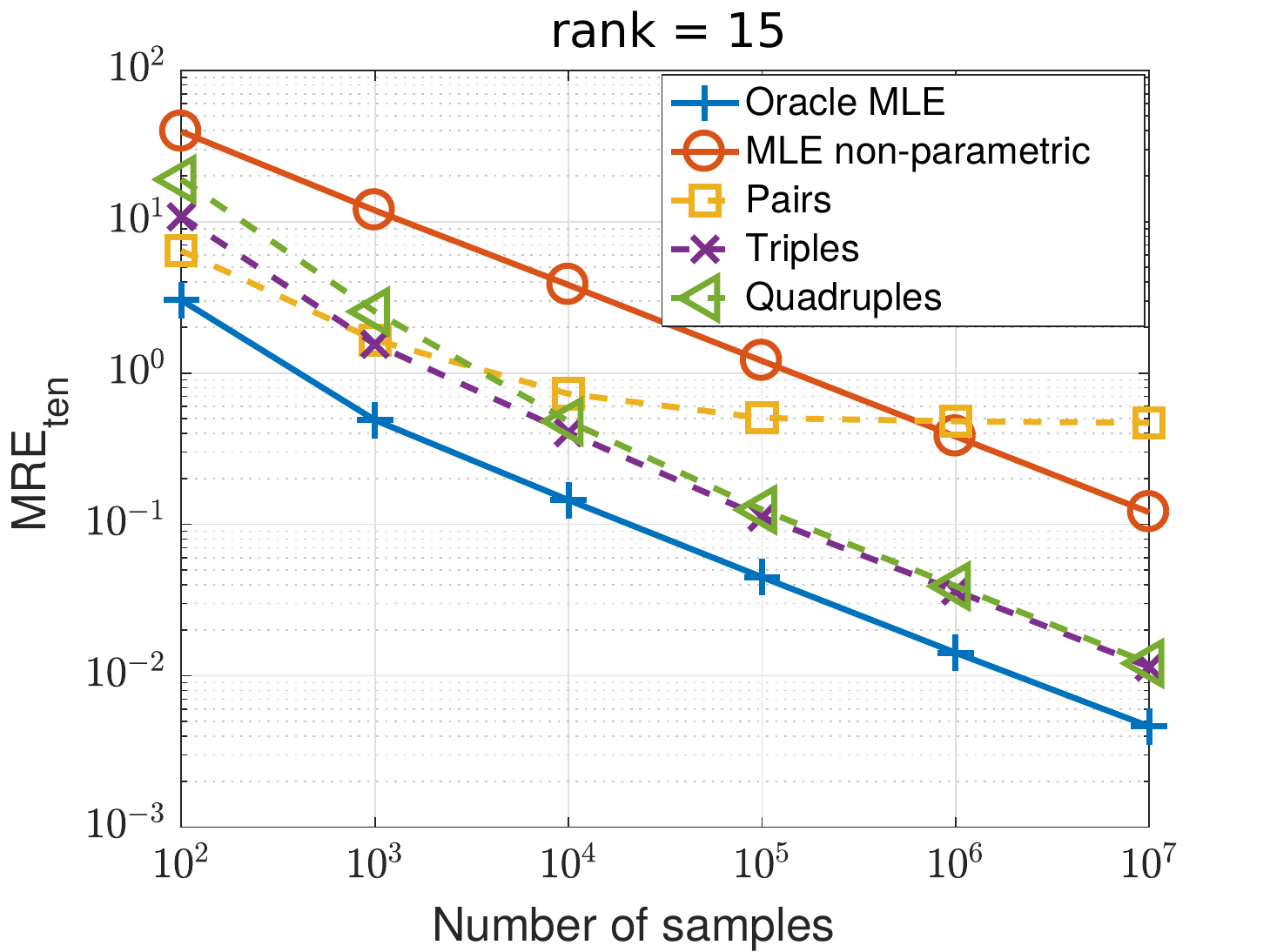}
\caption{Mean relative error of the estimated joint PMF under different number of available samples.}
\label{fig:synthetic_data2}
\end{figure*}

In this section, we employ judiciously designed synthetic data simulations to showcase the effectiveness of the proposed joint PMF recovery methods. We also apply the approach to real-data problems such as classification and recommender systems to demonstrate its usefulness in real machine learning tasks.

\begin{table}[!t]
\begin{center}
\caption{Mean relative factor and tensor error when  {lower-dimensional} marginals are perfectly known.}
\label{table:noiseless}
\begin{tabular}{c  l  c  c}
\hline
 Rank   &  & $\textrm{MRE}_{\textrm{fact}}$ & $\textrm{MRE}_{\textrm{ten}}$ \\
\hline
       & Pairs      &  $0.277$	              &  $0.148$                            \\
$F = 5$& Triples    &  $1.18 \times 10^{-7}$ &  $ 4.58 \times 10^{-8} $     \\
	   & Quadruples &  $3.39 \times 10^{-8}$ &  $ 1.19 \times 10^{-8}$		        \\
\hline
  		   &  Pairs       &  $ 0.440$			    &   $0.187$                      \\
$F = 10$   &  Triples     &  $3.58 \times 10^{-7}$	& 	$8.70 \times 10^{-8}$		       \\
	       &  Quadruples  &  $1.26 \times 10^{-7}$	& 	$2.58 \times 10^{-8}$		 \\
\hline
  		   &  Pairs       &  $0.466$			    &   $ 0.184$                      \\
$F = 15$   &  Triples     &  $6.77 \times 10^{-7}$	& 	$1.52 \times 10^{-7}$		       \\
	       &  Quadruples  &  $1.78 \times 10^{-7}$	& 	$3.57 \times 10^{-8}$\\
\hline
\end{tabular}
\end{center}
\end{table}
\subsection{Synthetic-Data Simulations}
We first evaluate the proposed approach using synthetic data. 
We consider a case where $N=5$ random variables are present, and each variable can take $I_n=10$ discrete values.
We assume that the joint PMF of the $5$ random variables can be represented by a naive Bayes model whose latent variable $H$ can take $F$ values, where $F$ is set to be $\{5,10,15\}$. 
We generate matrices $\mathbf{A}_n \in \mathbb{R}_+^{I_n \times F}$, that model the conditional probabilities i.e., ${\bf A}_n(i_n,f)={\sf Pr}(i_n|f)$.
A vector $\boldsymbol{\lambda} \in \mathbb{R}_+^F$ is also generated for the latent random variable $H$ such that ${\bm \lambda}(f)={\sf Pr}(f)$.
The elements of each ${\bf A}_n$ and the vector $\boldsymbol{\lambda}$ are drawn  { independently} from a uniform distribution between zero and one, and each column is normalized to sum to $1$. The ground-truth joint PMF is then constructed following the naive Bayes model, i.e., ${\sf Pr}(i_1,\ldots,i_5)=\underline{\mathbf{X}}(i_1,\ldots,i_5) = \sum_{f=1}^F \boldsymbol{\lambda}(f)  \prod_{n=1}^5 \mathbf{A}_n(i_n,f)$. We assume that the observable data are  {two-, three- and four-dimensional} marginals of the joint PMF.
Under such settings, we can verify if the proposed procedure and algorithm can effectively recover the joint PMF, if there is no modeling error and the joint PMF does have low rank.
We run $20$ Monte Carlo simulations and compute the mean relative error of the factors as well as the mean relative error of the recovered tensor which are defined as follows
\begin{equation*}
\textrm{MRE} _{\textrm{fact}}= \mathbb{E}\left( \frac{1}{N}  \sum_{n=1}^{N} \frac{ \| {\mathbf{A}}_n -  \widehat{{\mathbf{A}}}_n{\bm \Pi} \|_F}{ \| {\mathbf{A}}_n \|_F}\right),
\end{equation*}
\begin{equation*}
\textrm{MRE}_{\textrm{ten}} =  \mathbb{E}\left(\frac{ \| \underline{\mathbf{X}} - \widehat{\underline{\mathbf{X}}} \|_F}{ \| \underline{\mathbf{X}} \|_F}\right),
\end{equation*}
where ${\bm \Pi}$ is a permutation matrix to fix the permutation ambiguity,  and $\widehat{\underline{\mathbf{X}}}$,  $\widehat{{\mathbf{A}}}_n$ are the estimated joint PMF and the corresponding conditional PMFs.

Table~\ref{table:noiseless} shows the mean relative errors for estimating the conditional PMFs and joint PMF using the different types of input under different choices of rank. Consistent with our analysis, one can see that using marginal distributions of triples or quadruples of random variables (i.e.,  {three- and four-dimensional} marginals) we are able to recover the joint PMF of $5$ random variables. Here recovery with high accuracy has been demonstrated; exact recovery is also possible in certain cases, see \cite{SiDeFu2017}. However, using pairs (i.e.,  {two-dimensional} marginals) is not as promising. Recall that our identifiability result is built upon the identifiability of third- and higher-order CPD models. These nice identifiability results in general do not hold for matrices -- which explains the sharp performance difference between using the pairs and  {higher-dimensional} marginals. Pairs can work, however, when the conditional probability matrices are sufficiently sparse, and under more stringent constraints on the rank $F$. We defer detailed discussion to a follow up paper, due to lack of space in this one. 
 
The above simulation serves as sanity check -- if the available data and the model perfectly match with each other and we have noiseless marginal distributions, the proposed approach can indeed recover the joint PMF.
In practice, we usually do not have exact estimates of the  {lower-dimensional} marginal distributions.
Next, we provide a set of more realistic simulations where we estimate the marginal PMFs using sample averages from the observed data.

\subsubsection{Fully-observed data}

We follow the same way of generating the ground-truth joint PMF as in the previous simulation. Then, drawing from the joint PMF, we generate a synthetic dataset of $M$  {five-dimensional} data points. The data is generated as follows: for each data point, we first draw a sample $h_m$ according to $\boldsymbol{\lambda}$; i.e., a realization of the hidden variable $H$. Then the data point (vector)  { $\mathbf{s}_m = [\mathbf{s}_m(1),\ldots,\mathbf{s}_m(N)]^T$} is generated by drawing its elements independently from $\{\mathbf{A}_n\}(:,h_m)_{n=1}^N$, i.e., $\mathbf{s}_m(n)$ is drawn from $\{\mathbf{A}_n\}(:,h_m)$. This is equivalent to synthesizing the  {five-way} joint PMF tensor and drawing an outcome from it (cf. the naive Bayes interpretation). 

\begin{table*}[!ht]
\begin{center}
\caption{Misclassification error on different UCI datasets.}
\label{table:real_exp1}
\begin{tabular}{ l|  c  c  c c c | c c }
\hline
&  \multicolumn{5}{c}{\textbf{Binary}}           &  \multicolumn{2}{c}{\textbf{Multiclass}}       \\
\hline
 Method &  INCOME   &	CREDIT   &  HEART  	& MUSHROOM    &  VOTES & CAR & NURSERY  \\
\hline
 CP (Pairs) &  $0.177 \rpm 0.004$     & $ 0.134 \rpm 0.019$	 & $0.151 \rpm 0.023$ & $0.010 \rpm 0.007$  &      $0.046  \rpm 0.024$      & $0.128  \rpm 0.021 $  &$0.101 \rpm  0.009 $ \\
 CP (Triples)     &   $0.175 \rpm 0.003$      &  $ 0.129 \rpm 0.018 $ 	 & $0.147 \rpm 0.031 $   & $0.006 \rpm 0.002 $ &     $0.043 \rpm 0.024$      & $ 0.089 \rpm 0.016 $ &$ 0.069 \rpm 0.011 $ \\
 CP (Quadruples)  &  $\textbf{0.171} \rpm 0.003 $   &   $ \textbf{0.123} \rpm 0.018$  & $ \textbf{0.138}  \rpm 0.029$   & $0.002 \rpm 0.001$ &    $0.042 \rpm 0.020$       &  $ 0.074 \rpm 0.015 $  & $ 0.061 \rpm 0.007 $ \\
\hline
 SVM (Linear)     &	$0.179 \rpm 0.004$	    &  $0.146 \rpm 0.027$  &  $0.170 \rpm 0.053 $ & $ \textbf{0} \rpm 0$   &  $\textbf{0.038} \rpm 0.025$       & $0.065 \rpm 0.006$ & $0.063 \rpm 0.004$\\
 SVM (RBF)   	  & $0.174 \rpm 0.004$        & $0.136 \rpm 0.018$   & $0.187 \rpm 0.055$   & $\textbf{0} \rpm 0$ &    $0.079 \rpm 0.024$    & $ \textbf{0.026} \rpm 0.008$ & $ \textbf{0.006} \rpm 0.001$ \\
 Naive Bayes 	  & $0.209 \rpm 0.005$        &  $0.140 \rpm 0.018$  & $0.166 \rpm 0.026$  &  $0.044 \rpm 0.005 $   &  $0.096 \rpm 0.022$         &  $0.151 \rpm 0.016$  & $0.097 \rpm 0.007$ \\
\hline
\end{tabular}
\end{center}
\end{table*}

\begin{table}[!ht]
\begin{center}
\caption{Dataset information.}
\label{table:sparse_factors}
\begin{tabular}{l c c  c}
\hline
 Dataset    &  $ N $  & $F$  \\
\hline
 INCOME       & $8$ & $[1,20]$ \\
\hline
CREDIT     &  $9$ & $[1,20]$ \\
\hline
HEART      & $9$ & $[1,10]$ \\
\hline
MUSHROOM   &     $22$  & $[1,20]$ \\
\hline
VOTES      &  $17$  & $[1,10]$ \\
\hline
CAR        &  $7$  & $[1,15]$ \\
\hline
NURSERY     & $9$ &  $[1,15]$\\
\hline
\end{tabular}
\end{center}
\end{table}
We use the generated $5$-dimensional data points to estimate  {lower-dimensional} marginals and run our ADMM algorithm to recover the full joint PMF. We repeat for a total of $10$ Monte Carlo simulations. Figure~\ref{fig:synthetic_data} shows the tensor mean relative error of the estimated joint PMF under different dataset sizes $M$. We also include the performance of two additional methods for estimating the joint PMF. Given the full data together with `oracle' observations of the hidden variable $H$, we perform Maximum Likelihood Estimation (MLE) of the naive Bayes parameters, denoted as oracle MLE, which is done simply by frequency counting
\[
{\sf Pr}(f)_{ML} = \frac{{\rm count}(f)}{M}, \; {\sf Pr}(i_n| f)_{ML} = \frac{{\rm count}(i_n,f)}{{\rm count}(f)}, 
\]
where ${\rm count}(i_n,f)$ is the number of times that $X_n = i_n$ and $H = f$ appear together in the dataset and $ \rm count(f)$ the number of times $H$ takes the value $f$. We also include the MLE of a non-parametric approach in which we use the empirical $5$-dimensional distribution as our estimate i.e.,
\[
{\sf Pr}(i_1,\ldots,i_N)_{ML} = \frac{{\rm count}(i_1,\ldots,i_N)}{M}. \]
One can see that the estimation performance of our method is similar under different rank values and approaches that of oracle MLE using  {three- and four-dimensional} marginals. In addition, as expected, the recovery accuracy steadily improves as the size of the available dataset increases. On the other hand, when using  {two-dimensional} marginals the performance improves until it reaches a plateau at approximately $M = 10^5$. The ability to recover the joint PMF using pairs of random variables is obviously limited by identifiability of matrix factorization.

\subsubsection{Missing data}
We repeat the above experiment when some of the dataset entries are missing. We randomly hide $20\%$ of the data and compute estimates of  {two- three- and four-dimensional} marginals using only the available data. We run the ADMM-based algorithm and repeat for $10$ Monte Carlo simulations. The estimation performance of our method is again similar under different rank values and approaches that of oracle MLE. As expected, we observe a slight decrease in performance which is due to the less accurate estimation of the  {lower-dimensional} marginals. In this case, the MLE non-parametric method takes into account only the fully observed samples. 

 {Note that when empirical estimates of the lower-dimensional distributions are used and the number of samples is limited, three-dimensional distributions may give lower relative error compared to the four-dimensional ones. This shows that in some cases using lower-dimensional distributions can be more beneficial than higher-order ones in terms of parameter estimation accuracy. Actually, this is not very surprising since it can be shown that empirical lower-dimensional marginals are always more accurate than higher-dimensional ones when estimated given the same data~\cite{Huang2018}.}

\subsection{Real-Data Experiments}
In real applications, the ground-truth joint PMF and the conditional PMFs are not known. Nevertheless, we can evaluate the method on a variety of standard machine learning tasks to observe its effectiveness. 
In this subsection, we test the proposed approach on two different tasks, namely, data classification and recommender systems.
Note that both tasks can be easily accomplished if the joint PMF of pertinent variables (e.g., features and labels in classification) is known and thus are suitable for evaluating our method.
Note that the rank of the joint PMF tensor, or, $F=|{\cal H}|$, cannot be known as in the simulations. Fortunately, this is a single discrete variable that can be easily tuned, e.g., via observing validation errors as in machine learning.

\subsubsection{Classification Task}
We evaluate the performance of our approach on $7$ different datasets from the UCI machine learning repository~\cite{UCI}. Five of the selected datasets correspond to binary classification and two to multi-class classification. {For each dataset, we represent the training samples using its discrete features so that the PMF-based approach can be applied}. We split each dataset such that $70\%$ of the data samples is used for training, $10\%$ used for validation and $20\%$ for testing. 

\begin{table*}[!ht]
\begin{center}
\caption{RMSE and MAE of different algorithms on MovieLens (Ratings are in the range [1-5]).}
\label{table:regression}
\begin{tabular}{ l|  c  c | c c | c c }
\hline
&  \multicolumn{2}{c}{\textbf{MovieLens Dataset 1}}            &  \multicolumn{2}{c}{\textbf{MovieLens Dataset 2}}     &  \multicolumn{2}{c}{\textbf{MovieLens Dataset 3}} \\
\hline
 Method &  RMSE       & 	MAE     &  RMSE      			& MAE  &  RMSE      & MAE \\
\hline
 CP (Pairs)      & $ 0.802 \rpm 0.003 $   & 	$0.608 \rpm 0.003 $   	&  $0.795 \rpm 0.002 $ 	& $0.611 \rpm 0.002$  &  $ 0.897 \rpm 0.003$ & $0.702 \rpm 0.002 $ \\
 CP (Triples)    & $0.783 \rpm 0.002 $  & 	$0.591 \rpm 0.002$   	&  $\textbf{0.785 \rpm 0.002 }$  & $\textbf{ 0.599 \rpm  0.002}$  &  $ 0.887 \rpm 0.002 $ & $ 0.691 \rpm 0.002$  \\
 CP (Quadruples) & $\mathbf{0.778 \rpm 0.002}$    & 	$\mathbf{0.588 \rpm 0.002}$   &  $ 0.786 \rpm 0.002$ & $ 0.600 \rpm 0.002 $  &  $\textbf{0.884 \rpm 0.002}$       & $ \textbf{0.689 \rpm 0.002}$  \\
\hline
 Global Average   & $ 0.945 \rpm 0.001 $ & 	$0.693 \rpm 0.001 $     &  $ 0.906 \rpm 0.002$      & $ 0.653 \rpm 0.002 $ &  $0.996 \rpm 0.002 $   & $0.798 \rpm 0.001 $  \\
 User  Average    & $0.879 \rpm  0.002 $ & 	$ 0.679 \rpm 0.001$     &  $ 0.830 \rpm 0.003 $      & $ 0.625\rpm 0.002 $ &  $1.010 \rpm 0.002 $   & $ 0.768 \rpm 0.002$  \\
 Movie Average  	  & $0.886 \rpm 0.002$ & 	$0.705 \rpm 0.001 $  	&  $0.889 \rpm 0.002 $      & $ 0.673 \rpm 0.002$ &  $ 0.942 \rpm 0.002 $   & $ 0.754 \rpm 0.001$  \\
 BMF & $0.797 \rpm 0.002$ & 	$ 0.623 \rpm 0.002 $   	&  $ 0.792 \rpm 0.002 $ & $ 0.604 \rpm 0.002 $ &  $0.904 \rpm 0.003$   & $0.701 \rpm 0.003 $        \\
\hline
\end{tabular}
\end{center}
\end{table*}

For each dataset, we let $X_N$ be the label and $X_1,\ldots,X_{N-1}$ be the selected features. We estimate  {lower-dimensional} marginal distributions of pairs, triples and quadruples of variables using the samples in the training set. Then, we use the marginals to estimate ${\sf Pr}(i_n|f)$, and ${\sf Pr}(f)$.
After applying the proposed approach for estimating the joint PMF, we predict for each data point of the test set the corresponding label using the MAP rule. The MAP estimator of the label $l(\mathbf{s}_m)$ of the $m$-th observation $\mathbf{s}_m = [\mathbf{s}_m(1),\ldots,\mathbf{s}_m(N-1)]^T$ in the test set can be written as
\[
\widehat{l}_{\textrm{map}}(\mathbf{s}_m) = \argmax_{ i_N \in \{1,\ldots,I_N \} } {\sf Pr}( i_N | \mathbf{s}_m(1),\ldots,\mathbf{s}_m(N-1)),\]
where $I_N$ is the number of classes. Equivalently, using the Bayes rule the above can be found by
\[
\widehat{l}_{\textrm{map}}(\mathbf{s}_m) = \argmax_{ i_N \in \{1,\ldots,I_N \} } \sum_{f=1}^F {\sf Pr}(f) {\sf Pr}(i_N | f) \prod_{n=1}^{N-1} {\sf Pr}(\mathbf{s}_m(n)| f).
\]
For each dataset, we run $10$ Monte Carlo simulations {with randomly partitioned training/validation/test sets} and observe the average result. 
As mentioned, we do not know {\it a priori} what is an appropriate rank for our model. Therefore, for each dataset, we fit models of different rank values and choose the one which minimizes the misclassification error of the validation set as in standard machine learning practice.

We use $3$ different classical classifiers from the MATLAB Statistics and Machine Learning Toolbox  as baselines; linear SVM, kernel SVM with radial basis function and a naive Bayes classifier. For SVM classifiers, we use both the original data encoding as well as the one-hot encoding which usually is more suitable for discrete data and report the best result among the two.
Note that the baseline naive Bayes approach is very different from ours: the baseline method assumes that the features are independent given the label, while we assume that the label and the features are independent given an unknown latent variable.
The former is a very strong assumption that is rarely satisfied by real data, but our assumption holds for an arbitrary set of random variables provided $F$ is large enough, as we showed in Proposition~\ref{prop:F_upperbound}.

Table~\ref{table:real_exp1} shows the classification errors obtained on the datasets. One can see that our approach outperforms the naive Bayes classifier which assumes that the features are independent given the label. 
Several observations are in order.
First, using  {higher-dimensional} marginals, the proposed approach gives better classification results compared to using  {lower-dimensional} ones.
This is consistent with our analysis in Sec.~\ref{sec:ident} --  {higher-dimensional} marginals lead to stronger overall identifiability of the joint PMF. One can see that for all the datasets under test, using  {four-dimensional marginal} distributions gives the best classification accuracy compared to  { the three- and two-dimensional ones}.
Second, for the five binary classification experiments, the proposed method works better than (on three datasets) or comparable to the baselines. This is quite surprising since our method does not directly optimize a classification criterion as SVM does.
The result suggests that the proposed method indeed captures the essence of the joint distribution and the recovered joint PMF can be utilized to make inference in practice.
Third, for the multiclass datasets, the proposed method yields accuracy that is less than the SVM methods. This also makes sense: when $X_N$ has a value set whose cardinality grows from 2 to 5, the joint PMFs of $X_N$ and $X_i,X_j$ for $i,j<N$ require more samples to estimate accurately. This also shows an interesting sample complexity-accuracy trade-off of the proposed method.
Nevertheless, the method still works comparably well with the linear SVM, which supports the usefulness of the joint PMF estimation method.

\subsubsection{Recommender Systems}
We also evaluate the method for the task of recommender systems using the MovieLens dataset~\cite{HaMaKo2015}. MovieLens is a dataset that contains ratings on $5$-star scale, with half-star increments, by a number of users. In order to test our algorithm we select three different subsets of the full dataset and round the ratings to the next integer. Initially, three different categories (action, animation and romance) are selected. From each category we extract a user-by-rating submatrix by keeping the $20$ most rated movies and form the $3$ datasets for our experiments.  
Note that the constructed three datasets have many missing values, since not all users watched and rated all movies. The task of recommender systems is to recommend unwatched movies to users based on prediction of the user's rating given the available data.

We aim at estimating the joint PMF of the movie ratings. In this case, each random variable $X_n$ represents a movie, and it takes values from $\{1,\ldots,5\}$, i.e., the ratings.
Consequently, the joint PMF is a twenty-way tensor which has $5^{20}$ elements.
The $3$ partially observed datasets are used in order to estimate  {lower-dimensional} marginal distributions of pairs, triples and quadruples of the variables (movies). We use the estimated PMF to compute the expected value of  users' ratings that we do not observe given the ones we observe. More specifically, let  $\mathbf{s}_m = [\mathbf{s}_m(1),\ldots,\mathbf{s}_m(N)]$ be the ratings of the $m$-th user and $\mathbf{s}_m(N)=0$ i.e., the user has a missing rating. The conditional expectation of the movie's rating is given by
\[
\widehat{s}_N = \sum_{i_N = 1}^{I_N} i_N {\sf Pr}(i_N | \mathbf{s}_m(1),\ldots,\mathbf{s}_m(N-1)).
\]
As a baseline algorithm, we use the Biased Matrix Factorization (BMF) method~\cite{Koren2009}, which is a commonly used method in recommender systems. The BMF method is essentially low-rank matrix completion with modifications.
Additionally, we present results  obtained by global average of the ratings, the user average, and the item average  as baselines for predicting the missing entries. For each dataset we randomly hide $20\%$ ratings that we use as a test set, $10\%$ ratings that we use as a validation set and the remaining dataset is used as a training set. We run $10$ Monte Carlo simulations using our approach and the BMF algorithm. We select the parameters of both methods based on the RMSE of the validation set. 

Table~\ref{table:regression} shows the performance of the two algorithms in terms of the RMSE and Mean Absolute Error (MAE). One can see that, for the three datasets under test, the proposed method and the BMF method output clearly lower RMSEs and MAEs relative to the naive methods using averaging. In addition, the proposed method slightly outperforms BMF on all of the three datasets. Note that BMF is considered a state-of-art method for movie recommendation, and it incorporates application-specific custom features, such as user bias and movie bias to achieve good performance. The proposed method, on the other hand, only uses basic probability to handle the same task -- it is completely application-blind. This suggests that the joint PMF modeling and the proposed algorithm are both quite effective. Last, we also observe accuracy improvement when we increase the dimension of the marginal distributions used in the approach. Again, this performance may come from the identifiability gain as we analyzed in Theorem~\ref{thm:proposed_3}.

\section{Conclusions}
\label{sec:conclusion}

In this work, we have taken a fresh look at one of the most fundamental problems in statistical learning -- joint PMF estimation. Due to the curse of dimensionality, naive count-based estimation is mission impossible in most cases. One popular approach has historically been to assume a plausible structural model, such as a Markov chain, tree, or other probabilistic graphical model, and do inference using this model. We showed that a very different `non-parametric' tensor-based approach is possible, and it features several key benefits. Foremost among them is guaranteed identifiability of the  {high-dimensional} joint PMF from  {low-dimensional marginals}, which can be reliably estimated using counting from much fewer examples, even if there are (many) samples missing from each example. This ability to infer a unique  {higher-dimensional} joint PMF by specifying  {lower-dimensional} ones is reminiscent of Kolmogorov extension, which is intuitively very pleasing. 

We have also proven two more results that appear fundamental and close to the heart of probability and learning theory: i) {\em every} joint PMF can be {\em interpreted} as a naive Bayes model; and ii) probabilistic graphical models, which are very popular in statistical and computer science, are never identifiable if one simply limits the number of hidden nodes; one needs to bound the number of hidden states as well. 
Our non-parametric approach can reveal the true hidden structure, instead of assuming it; and this alleviates the risk of up-front bias in the analysis. 

On the practical side, our approach is appealing since  {lower-dimensional} marginals can be more reliably estimated from a limited amount of partially observable data. We have also provided a practical and easily implementable algorithm that is based  on factor-coupled tensor factorization to handle the recovery problem. Simulations and judicious experiments with real data have shown that the performance of the proposed approach is consistent with our analysis, and approaches or exceeds that of state-of-art application-specific solutions that have come out after years of intensive research, which is satisfying. 

\begin{appendices}
\section{Identifiability Results}
\label{ap:ident_results}

\subsection{Proof of Theorem~\ref{thm:proposed_1}}

Each  {three-dimensional marginal satisfies} $\underline{\mathbf{X}}_{jkl} = [\![ \boldsymbol{\lambda},\mathbf{A}_j,\mathbf{A}_k,\mathbf{A}_l ]\!]$, where $\mathbf{A}_l \geq \mathbf{0}$, $\mathbf{A}_k \geq \mathbf{0}$, $\mathbf{A}_j \geq \mathbf{0}$, $\mathbf{1}^T \mathbf{A}_l = \mathbf{1}^T$, $\mathbf{1}^T \mathbf{A}_k = \mathbf{1}^T$, $\mathbf{1}^T \mathbf{A}_j = \mathbf{1}^T$, $\boldsymbol{\lambda} > \mathbf{0} $, $\mathbf{1}^T \boldsymbol{\lambda} = 1$.
Consider a partition of the set $\mathcal{S} = [N]$ into three disjoint sets $\mathcal{S}_1,\mathcal{S}_2,\mathcal{S}_3$ and define the following factors
\begin{equation}
\begin{aligned}
	\widehat{\mathbf{A}}_1 &=  [\mathbf{A}_{u_1}^T, \cdots, \mathbf{A}_{u_{|\mathcal{S}_1|}}^T ]^T  \\
	\widehat{\mathbf{A}}_2 &= [ \mathbf{A}_{v_1}^T, \cdots, \mathbf{A}_{v_{|\mathcal{S}_2|}}^T ]^T \\
	\widehat{\mathbf{A}}_3 &= [ \mathbf{A}_{w_1}^T, \cdots,\mathbf{A}_{w_{|\mathcal{S}_3|}}^T ]^T
\end{aligned}
\end{equation}
with $u_t \in \mathcal{S}_1$, $v_t \in \mathcal{S}_2$, $w_t \in \mathcal{S}_3$. Then, we can construct a  single virtual nonnegative CPD model
\begin{equation}
\underline{\widehat{\mathbf{X}}}^{(1)} = (\widehat{\mathbf{A}}_3 \odot \widehat{\mathbf{A}}_2) \textrm{diag}(\boldsymbol{\lambda})\widehat{\mathbf{A}}_1^T,
\end{equation} 
where $\widehat{\mathbf{A}}_1 \in \mathbb{R}_+^{I|\mathcal{S}_1| \times F}, \widehat{\mathbf{A}}_2 \in \mathbb{R}_+^{I|\mathcal{S}_2| \times F}, \widehat{\mathbf{A}}_3 \in \mathbb{R}_+^{I|\mathcal{S}_3| \times F}$ and $ \underline{\widehat{\mathbf{X}}} \in \mathbb{R}_+^{I |\mathcal{S}_1| \times I |\mathcal{S}_2| \times I |\mathcal{S}_3|}$.
We have used a subset of the available information to synthesize a virtual single nonnegative CPD model of size $I_1 \times I_2 \times I_3$, with $I_k := I|\mathcal{S}_k|$. 
Therefore, we can apply identifiability results of  {three-way} tensors. We observe that the sizes of the different modes of the virtual tensor depend on the way we partition the variables. We distinguish between two cases and apply Lemma~\ref{lem:generic_1}.

\begin{enumerate}[leftmargin=*]
\item  { $ {(N \leq I)}$}: We partition the variables into three sets such that $I_1 = I, I_2 = I$ and $I_3 = (N-2)I$. Clearly we have that $I_3 > I_2,I_1 $ and $I_3 < (I_1-1)(I_2-1)$. According to Lemma~\ref{lem:generic_1} tensor $\underline{\widehat{\mathbf{X}}}$ admits unique decomposition 
for $F\leq \min \left(I_3,(I_1-1)(I_2-1) \right) = (N-2)I$.
\item  { $ {(N>I)}$}: In this case we can partition the variables into three sets such that $I_3 = (I_1-1)(I_2-1)$. We can always have that $I_1 = I_2$. The equality is satisfied when $ |\mathcal{S}_1| = |\mathcal{S}_2|= \frac{\sqrt{(NI-1)}}{I}$. According to Lemma~\ref{lem:generic_1} tensor $\underline{\widehat{\mathbf{X}}}$ admits unique decomposition for $F\leq \min (I_3,(I_1-1)(I_2-1)) \leq ( \lfloor \frac{\sqrt{(NI-1)}}{I}\rfloor I -1)^2 $.
\end{enumerate}

\subsection{Proof of Theorem~\ref{thm:proposed_2}}
\label{ap:ident_results1}
As above, but this time choosing $|\mathcal{S}_1| = |\mathcal{S}_2| = \lfloor \frac{N}{3} \rfloor$,  {$|S3| = N - 2|S1|$} and invoking Lemma~\ref{lem:generic_2}.

\subsection{Proof of Theorem~\ref{thm:proposed_3}}
\label{ap:ident_results2}

Consider a partitioning the variables into four disjoint sets similar to the  {three-way} case. We obtain a single nonnegative CPD of the following form
\begin{equation*}
\underline{\mathbf{X}}^{(1)} = (\widehat{\mathbf{A}}_4 \odot \widehat{\mathbf{A}}_3 \odot \widehat{\mathbf{A}}_2) \textrm{diag}(\boldsymbol{\lambda})\widehat{\mathbf{A}}_1^T,
\end{equation*} 
which is a fourth-order tensor $\underline{\widehat{\mathbf{X}}} \in \mathbb{R}_+^{I |\mathcal{S}_1| \times I |\mathcal{S}_2| \times I |\mathcal{S}_3| \times I |\mathcal{S}_4|}$. A fourth-order tensor can be viewed as a third-order tensor of size $I|\mathcal{S}_1|\times I|\mathcal{S}_2|\times I^2|\mathcal{S}_4||\mathcal{S}_3|$ with a specially structured factor matrix. We can write the mode-$1$ unfolding of the tensor $\underline{\widehat{\mathbf{X}}}$ as
\begin{equation}
\underline{\widehat{ \mathbf{X}} }^{(1)} = ( \bar{\mathbf{A}}_3 \odot \widehat{\mathbf{A}}_2) \textrm{diag}(\boldsymbol{\lambda})\widehat{\mathbf{A}}_1^T,
\label{eq:4_way}
\end{equation} 
where $\bar{\mathbf{A}}_3 = \widehat{\mathbf{A}}_4 \odot \widehat{\mathbf{A}}_3 $.
Lemmas~\ref{lem:generic_1}-\ref{lem:generic_2} cannot be applied in this case because of the Khatri-Rao structure of one factor. We will use the following result
\begin{Lemma}~\cite{JiSi2004} \label{thm:jisi2004} Let $\underline{\mathbf{X}} = [\![ \mathbf{A}_1,\mathbf{A}_2,\mathbf{A}_3 ]\!]$, where $\mathbf{A}_1 \in \mathbb{R}^{I_1 \times F}$, $\mathbf{A}_2 \in \mathbb{R}^{I_2 \times F}$, $\mathbf{A}_3 \in \mathbb{R}^{I_3 \times F}$. If $\textrm{rank}(\mathbf{A}_3)=F$ and $\textrm{rank} \big(\mathbf{M}_2(\mathbf{A}_1) \odot \mathbf{M}_2(\mathbf{A}_2) \big ) = \binom{F}{2}$, then $\textrm{rank}(\underline{\mathbf{X}})$ = $F$ and the decomposition is essentially unique.
\end{Lemma}

$\mathbf{M}_2(\mathbf{A})$ denotes the $\binom{I}{2}\times\binom{F}{2}$ compound matrix containing all $2\times 2$ minors of $\mathbf{A}$. The generic version of Lemma~\ref{thm:jisi2004} states that if $F \leq I_3$ and $2F(F-1) \leq I_1 (I_1 -1)I_2(I_2-1)$, then the decomposition of $\underline{\mathbf{X}}$ is essentially unique, a.s.~\cite{SteBeLa2006}. We know that a Khatri-Rao product of two matrices is full rank almost surely~\cite[Corollary 1]{JiSi2001}. For the {three-way} tensor in~\eqref{eq:4_way} we have $I_1 = I |\mathcal{S}_1|$, $I_2 = I |\mathcal{S}_2|$ and $I_3 = I^2 |\mathcal{S}_3| |\mathcal{S}_4|$. Applying the generic version of Lemma~\ref{thm:jisi2004} we obtain the desired result.

\section{Algorithm}

\label{ap:algorithm}
We reformulate optimization problem~\eqref{opt:coupled_tensor_sub1} by introducing  an auxiliary variable $\widehat{\mathbf{A}}_j$. The problem can be equivalently written as
\begin{equation}
\begin{aligned}
&  \min_{ \mathbf{A}_j,\widehat{\mathbf{A}}_j}  f(\widehat{\mathbf{A}}_j) + r(\mathbf{A}_j) \\
& \text{subject to} \; \mathbf{A}_j = \widehat{\mathbf{A}}_j^T
\end{aligned}
\label{opt:equiv}
\end{equation}
where 
\begin{equation*} 
f(\widehat{\mathbf{A}}_j) = \sum_{k \neq j} \sum_{\substack{l\neq j \\ l > k}}  \frac{1}{2} \left \| \mathbf{X}_{jkl}^{(1)} - (\mathbf{A}_l \odot \mathbf{A}_k) \textrm{diag}(\boldsymbol{\lambda}) \widehat{\mathbf{A}}_j \right \|_F^2 ,
\end{equation*}
$r(\mathbf{A})$ is the indicator function for the probability simplex constraints $\mathcal{C} = \{ \mathbf{A}  \mid \mathbf{A}\geq 0, \mathbf{1}^T\mathbf{A} = \mathbf{1}^T\}$
\begin{equation*}
r(\mathbf{A}) = \begin{cases}
0, \;\;\;  \mathbf{A} \in \mathcal{C}\\
\infty,  \; \mathbf{A} \notin \mathcal{C}
\end{cases}.
\end{equation*}
Optimization problem~\eqref{opt:equiv} can  be readily solved by applying the ADMM algorithm.  We solve for $\mathbf{A}_j$ by performing the following updates
\begin{equation*}
\begin{aligned}
& \widehat{\mathbf{A}}_j^{(\tau + 1)} = \left( \mathbf{G}_j + \rho \mathbf{I} \right )^{-1} \left( \mathbf{V}_j  + \rho \left ( \mathbf{A}_j^{(\tau)} + \mathbf{U}_j^{(\tau)} \right )^T \right), \\
& \mathbf{A}_j^{(\tau + 1)}  =  \mathcal{P}_{\mathcal{C}}\left ( \mathbf{A}_j^{(\tau)} - \widehat{\mathbf{A}}_j^{(\tau + 1)T} + \mathbf{U}_j^{(\tau)} \right), \\
& \mathbf{U}_j^{(\tau + 1)} = \mathbf{U}_j^{(\tau)} + \mathbf{A}_j^{(\tau + 1)} - \widehat{\mathbf{A}}_j^{(\tau + 1)T}, \\
\end{aligned}
\end{equation*}
where
\begin{equation*}
\begin{aligned}
& \mathbf{G}_j = (\boldsymbol{\lambda}\boldsymbol{\lambda}^T) \circledast  \sum_{k \neq j}  \sum_{\substack{l\neq j \\ l > k}}  \mathbf{Q}_{lk}^T\mathbf{Q}_{lk}, \\
& \mathbf{V}_j = \textrm{diag}(\boldsymbol{\lambda}) \sum_{k \neq j}  \sum_{\substack{l\neq j \\ l > k}} \mathbf{Q}_{lk}^T {\mathbf{X}}_{jkl}^{(1)}, \\
& \mathbf{Q}_{lk} = \mathbf{A}_l \odot \mathbf{A}_k.
\end{aligned}
\end{equation*}
$\mathcal{P}_{\mathcal{C}}$ is the projection operator onto the convex set  $\mathcal{C}$ and $\tau$ denotes the iteration index. Various methods exist for projecting onto the probability simplex, e.g., see~\cite{WaCa2013}. Note that in order to efficiently compute matrix $\mathbf{G}_j$ we use a property of the Khatri-Rao product; $\mathbf{Q}_{lk}^T\mathbf{Q}_{lk} = (\mathbf{A}_l^T\mathbf{A}_l )\circledast(\mathbf{A}_k^T\mathbf{A}_k)$. Efficient algorithms also exist for the computation of matrix $\mathbf{V}_j$ which is a sum of Matricized Tensor Times Khatri-Rao Product (MTTKRP) terms~\cite{BaKo2008,SmiRaSi2015}. 
Similarly, we can derive updates for $\boldsymbol{\lambda}$.  At each ADMM iteration we perform the following updates
\begin{equation*}
\begin{aligned}
\widehat{\boldsymbol{\lambda}}^{(\tau+1)} & =  (\mathbf{G} + \rho \mathbf{I})^{-1} \left( \mathbf{V}  + \rho \left ( \boldsymbol{\lambda}^{(\tau)} + \mathbf{u}^{(\tau)} \right) \right), \\
\boldsymbol{\lambda}^{(\tau + 1)} & =  \mathcal{P}_{\mathcal{C}} \left( \boldsymbol{\lambda}^{(\tau)} - \widehat{\boldsymbol{\lambda}}^{(\tau + 1)} + \mathbf{u}^{(\tau)} \right), \\
\mathbf{u}^{(\tau + 1)} & =  \mathbf{u}^{(\tau)} + \boldsymbol{\lambda}^{(\tau+1)} - \widehat{\boldsymbol{\lambda}}^{(\tau + 1)}.
\end{aligned}
\end{equation*}
where
\begin{equation*}
\begin{aligned}
& \mathbf{G} =  \sum_j \sum_{ k > j} \sum_{l>k}  \mathbf{Q}_{lkj}^T\mathbf{Q}_{lkj}, \\
& \mathbf{V} = \sum_j \sum_{ k > j} \sum_{l>k} \mathbf{Q}_{lkj}^T \text{vec}(\underline{\mathbf{X}}_{jkl}), \\
& \mathbf{Q}_{lkj} = \mathbf{A}_l \odot \mathbf{A}_k \odot \mathbf{A}_j.
\end{aligned}
\end{equation*}

\end{appendices}

\bibliographystyle{IEEEtran}

\bibliography{IEEEabrv,lib}

\end{document}